\documentclass[prd,10pt,aps,twocolumn,superscriptaddress,floatfix,notitlepage,nofootinbib,amssymb,amsmath]{revtex4-1}
\usepackage{mathtools}
\usepackage{epsfig}
\usepackage{color}
 
\newcommand{\beqn}{\begin{eqnarray}}
\newcommand{\eeqn}{\end{eqnarray}}
\newcommand{\eq}[1]{(\ref{#1})}

\newcommand{\cO}{{\cal O}}

\newcommand{\cL}{{\cal L}}
\newcommand{\cZ}{{\cal Z}}

\newcommand{\cA}{{\cal A}}
\newcommand{\cD}{{\cal D}}

\newcommand{\ph}{{\mathrm {ph}}}
\newcommand{\mon}{{\mathrm {mon}}}

\newcommand{\lat}{{\mathrm{lat}\,}}
\newcommand{\bx}{\boldsymbol {x}}
\newcommand{\by}{\boldsymbol {y}}
\newcommand{\vx}{\vec {x}}
\newcommand{\vy}{\vec {y}}
\newcommand{\vq}{\vec {q}}
\newcommand{\Cas}{{\mathrm{Cas}\,}}
\newcommand{\Z}{{\mathbb Z}}
\newcommand{\bs}{\boldsymbol}
\newcommand{\plane}{{\cal P}_{\cal S}}
\newcommand{\avr}[1]{{\left\langle #1 \right\rangle}}
\newcommand{\aavr}[1]{\avr{\!\avr{ #1 }\!}}

\newcommand{\sumint}{{\mathclap{\displaystyle\int}\mathclap{\textstyle\sum}}}

\begin{document}

\title{Nonperturbative Casimir effect and monopoles: \\
compact Abelian gauge theory in two spatial dimensions}

\author{M. N. Chernodub}
\affiliation{Laboratoire de Math\'ematiques et Physique Th\'eorique UMR 7350, Universit\'e de Tours, 37200 France}
\affiliation{Laboratory of Physics of Living Matter, Far Eastern Federal University, Sukhanova 8, Vladivostok, 690950, Russia}
\author{V. A. Goy}
\affiliation{Laboratory of Physics of Living Matter, Far Eastern Federal University, Sukhanova 8, Vladivostok, 690950, Russia}
\author{A. V. Molochkov}
\affiliation{Laboratory of Physics of Living Matter, Far Eastern Federal University, Sukhanova 8, Vladivostok, 690950, Russia}

\begin{abstract}
We demonstrate that Casimir forces associated with zero-point fluctuations of quantum vacuum may be substantially affected by the presence of dynamical topological defects. In order to illustrate this nonperturbative effect we study the Casimir interactions between dielectric wires in a compact formulation of Abelian gauge theory in two spatial dimensions. The model possesses topological defects, instanton-like monopoles, which are known to be responsible for nonperturbative generation of a mass gap and for a linear confinement of electrically charged probes. Despite the model has no matter fields, the Casimir energy depends on the value of the gauge coupling constant. We show, both analytically and numerically, that in the strong coupling regime the Abelian monopoles make the Casimir forces short-ranged. Simultaneously, their presence increases the interaction strength between the wires at short distances for certain range of values of the gauge coupling. The wires suppress monopole density in the space between them compared to the density outside the wires. In the weak coupling regime the monopoles become dilute and the Casimir potential reduces to a known theoretical result which does not depend on the gauge coupling. 
\end{abstract}

\date{March 6, 2017}

\maketitle

\section{Introduction}

The Casimir effect is associated with a force which emerges between neutral objects  due to vacuum fluctuations of quantum fields~\cite{ref:Casimir,ref:Bogdag,ref:Milton}. The presence of the physical objects affects the spectrum of quantum fluctuations around them, and, therefore,  leads to a modification in the (vacuum) energy of these quantum fluctuations. Since the latter depends on mutual positions and orientations of the objects, the fluctuations of the vacuum fields naturally give rise to a force (called sometimes Casimir-Polder force) which acts on these objects~\cite{Casmir:1947hx}.

The simplest and best known example of the Casimir effect is given by interaction of two neutral parallel plates in the vacuum. The zero-point fluctuations of electromagnetic field give rise to attraction of idealized perfectly conducting plates. If the plates are made of real materials and/or are immersed in a medium, then the force between the plates becomes dependent on intrinsic properties of the corresponding materials such as permittivity, permeability and conductivity~\cite{Lifshitz:1956zz}. In certain cases the Casimir forces can even be made repulsive~\cite{ref:repulsive:Casimir} which is of immense practical interest for assembling frictionless (nano)mechanical machines.

Even in free noninteracting theories the calculation of the Casimir forces between objects of general shapes is a difficult task. There are not so many geometrical configurations for which the Casimir energy is known analytically. In order to compute of the Casimir interactions between objects of arbitrary geometries, and in the case if they are made of imperfect materials, a set of dedicated numerical methods has been developed~\cite{Johnson:2010ug,Gies:2006cq}. 

The zero-point forces are generally affected by interactions between quantum fields. In the case of Quantum Electrodynamics the one-loop perturbative correction to the Casimir effect is extremely small so that is can safely be neglected~\cite{ref:Milton}. On the other hand, in the fermionic version of the Casimir effect the four-point interaction modifies significantly the Casimir force~\cite{Flachi:2013bc}. In this case the virtual fermions of the vacuum are confined in between ``reflective'' plates with the MIT boundary conditions. The constrained fermionic excitations lead to appearance of the vacuum forces which are acting on the plates. On the other hand, the presence of the plates affects the structure of the fermionic vacuum which, in this model, may exist in the phases with spontaneously broken and restored chiral symmetry. At small separations between the plates the spontaneously broken phase ceases to exist~\cite{Flachi:2013bc,Tiburzi:2013vza}. A similar effect appears also in a cylindrical geometry which may by further enforced by a uniform rotation of the cylinder around its axis~\cite{Chernodub:2016kxh}.

In Ref.~\cite{Chernodub:2016owp} we proposed a general numerical method to study Casimir forces using first-principle simulations of lattice (gauge) theories. These methods, which are widely used in particle physics, are especially convenient for investigation of interacting theories as well as for studying nonperturbative features such as, for example, the phase structure of a given theory in a finite Casimir geometry or nonperturbative corrections to the Casimir force. In order to demonstrate the reliability of the method, in Ref.~\cite{Chernodub:2016owp} we have computed the Casimir energy between thin dielectric wires of finite permittivity in an Abelian gauge theory in two spatial dimensions. We have also shown that in the special case of straight wires of infinite permittivity our approach gives rise to a known analytic result. A similar numerical method for the case of ideal conductors has also been developed in Ref.~\cite{ref:Oleg}. 

The method of Ref.~\cite{Chernodub:2016owp} allows us to address straightforwardly nonperturbative properties of the gauge fields. There are two interesting effects associated with nonperturbative dynamics of the gauge fields, these are charge confinement and mass gap generation. Both these phenomena exist in non-Abelian gauge theories, of which is of particular interest is Quantum Chromodynamics. A useful toy Abelian model which possesses both these effects is the compact U(1) gauge theory in two space dimensions.  Although there is no matter fields in the theory, it is often called ``compact electrodynamics''  (compact QED) for shortness. We will use this terminology below. Before proceeding further we would like to notice that the compact QED serves as a toy model not only for particle physics, but it is also a viable effective model in a number of condensed matter applications~\cite{ref:Herbut}.

The compactness of the Abelian group leads to the appearance of certain topological objects (often called ``topological defects'') which carry magnetic charges and thus are associated with monopoles. Contrary to the usual particles, in two space time the world-trajectories of these magnetic defects are points, so that the monopoles are, in fact, instantons. At zero temperature these monopoles for a gas, which lead to both nonperturbative properties: to the linear confinement and to the mass gap generation. The confinement reveals itself in the total energy of a pair of particles, which carry positive and negative unit of electric charge. The energy of the pair rises linearly with the distance at large enough separations of the constituents of the pair. The particles in the pair are thus confined since in order to separate them to infinite distance one would need infinite energy. The mass gap generation is revealed via a finite mass of the photon. Both the confining force and the mass of the photon are proportional to the square root of the monopole density, and they cannot be derived in a standard perturbation theory in powers of electric coupling of the model~\cite{Polyakov:1976fu}.

The aim of this paper is to study nonperturbative effects of the monopoles on the Casimir forces between two parallel wires in the vacuum of compact electrodynamics. In addition, we also investigate how the vacuum affects the phase structure of the theory. We concentrate our efforts on zero-temperature case leaving investigation of the finite-temperature effects for a future. 

The structure of the paper is as follows. We review compact QED in continuum spacetime in Sect.~\ref{sec:cQED:review}. Then we calculate the influence of the monopole on Casimir energy in the dilute gas approximation in Sect.~\ref{sec:compact:QED:monopoles}. The nonperturbative numerical calculations of the Casimir effect are in the lattice formulation of the model in Sect.~\ref{sec:compact:QED:monopoles}. We summarize our conclusions in the last section.

\section{(2+1) compact electrodynamics}
\label{sec:cQED:review}

In this section we review in details certain well-known features of compact electrodynamics -- i.e. an Abelian theory with compact gauge group -- in two spatial dimensions in continuum spacetime. These properties will later be relevant to our studies of the nonperturbative Casimir effect both in the continuum spacetime (Sect.~\ref{sec:compact:QED:monopoles}) and in the lattice formulation of the model (Sect.~\ref{sec:compact:lattice:QED}).

\subsection{Monopoles and photons}

\subsubsection{Lagrangian of (2+1) compact electrodynamics}

The compact electrodynamics is basically a pure U(1) gauge theory with monopoles. In (3+1) dimensions the monopoles are particle-like objects and their world trajectories are closed lines that share similarly with usual pointlike particles. In (2+1) dimensions the ``trajectories'' of the monopoles are represented as a set of localized points, so that the monopoles are instanton-like objects. We will consider the (2+1) dimensional compact electrodynamics where most calculations can be done analytically. Since we are working in the thermal equilibrium and study stationary phenomena, it is convenient to perform a Wick rotation and consider  the compact electrodynamics in three-dimensional Euclidean spacetime. In this section we follow Ref.~\cite{Polyakov:1976fu}.

The Lagrangian of the compact electrodynamics has the same form as the Lagrangian of the usual free U(1) gauge theory:
\beqn
\cL = \frac{1}{4} F^2_{\mu\nu}\,,
\label{eq:L}
\eeqn
where $F_{\mu\nu}$ is the field strength tensor. However, contrary to a free U(1) gauge theory, the field strength tensor consists of two parts:
\beqn
F_{\mu\nu} = F_{\mu\nu}^{\ph} + F_{\mu\nu}^{\mon}\,.
\label{eq:F:munu}
\eeqn
The first, perturbative term is expressed via the vector photon field $A_\mu$,
\beqn
F_{\mu\nu}^{\ph}[A] = \partial_\mu A_\nu - \partial_\nu A_\mu\,,
\label{eq:F:ph}
\eeqn
while the second, nonperturbative monopole part,
\beqn
F_{\mu\nu}^{\mon}(\bx) = - g_\mon \epsilon_{\mu\nu\alpha} \partial_\alpha \int d^3 y D(\bx-\by) \rho(\by)\,,
\label{eq:F:mon}
\eeqn
is determined by the density of the instanton-like monopoles:
\beqn
\rho(\bx) = \sum_a q_a \delta^{(3)} \left(\bx - \bx_a\right)\,.
\label{eq:rho}
\eeqn 
Here ${\bs x}_a$ is the position of the $a$-th monopole, $q_a$ is its charge in units of the usual, Dirac-quantized elementary monopole charge,\footnote{
Here we use the standard Dirac quantization~\eq{eq:g:m} as compared to the quantization $g g_\mon/(4 \pi ) \in \Z$ which is the more appropriate for the monopole charges of the 't~Hooft-Polyakov monopoles in non-Abelian, grand unified theories~\cite{tHooft:1974kcl,Polyakov:1974ek}.}
\beqn
g_\mon = \frac{2 \pi}{g}\,,
\label{eq:g:m}
\eeqn
which, in turn, is expressed via the elementary electric charge $g$. In three-dimensional Euclidean spacetime the Greek indices run through $\mu,\nu,\alpha, \ldots = 1,2,3$. Notice that in (2+1) Minkowski (or, equivalently) 3 Euclidean dimensions the electric charge is a dimensional quantity, $[g] = {\mathrm{mass}}^{1/2}$, so that the monopole charge is also a dimensionful quantity, $[g_\mon] = {\mathrm{mass}}^{-1/2}$.

The monopole field strength tensor~\eq{eq:F:mon} depends nonlocally on the monopole density $\rho(\bx)$. The quantity $D$ in Eq.~\eq{eq:F:mon} is the scalar propagator,
\beqn
D({\bs x}) = \int \frac{d^3 k}{(2 \pi)^3} \frac{e^{i {\bs k} \bx}}{k^2} = \frac{1}{4 \pi | \bx |}\,,
\label{eq:D}
\eeqn
which obeys the second-order differential equation:
\beqn
- \Delta D(\bx) = \delta(\bx)\,,
\eeqn
where $\Delta \equiv \partial_\mu^2$ is the three-dimensional Laplacian.

The action of the model,
\beqn
S [A,\rho] = \frac{1}{4}  \int d^3 x \, F^2_{\mu\nu}
\label{eq:S:continuum}
\eeqn
decouples into a sum
\beqn
S [A,\rho] & = & \frac{1}{4} \int d^3 x \, \bigl(F^\ph_{\mu\nu}[A] + F^\mon_{\mu\nu}[\rho] \bigr)^2 \nonumber \\
& \equiv &
S^\ph[A] + S^\mon[\rho]\,, 
\label{eq:S:A:rho}
\eeqn
of the perturbative photon part
\beqn
S^\ph[A] & = & \frac{1}{4} \int d^3 x \, \bigl(F^\ph_{\mu\nu}[A]\bigr)^2\,, 
\label{eq:S:A}
\eeqn
and monopole part
\beqn
S^\mon[\rho] & = & \frac{g^2_\mon}{2} \int d^3 x \, \int d^3 y \, \rho({\bs x}) D(x-y) \rho({\bs y}), \quad
\label{eq:S:rho}
\eeqn
where we have implemented integration by parts and used the explicit forms of the photon and monopole field strengths given in Eqs.~\eq{eq:F:ph} and \eq{eq:F:mon}, respectively. The photon-monopole cross-term disappears due to the identity coming from the explicit form of the monopole field strength~\eq{eq:F:mon}:
\beqn
\partial_\mu F^{\mu\nu}_\mon \equiv 0\,.
\label{eq:dF:1}
\eeqn

Using the explicit expression for the monopole density~\eq{eq:rho} one can rewrite the monopole action~\eq{eq:S:rho} in terms of the Coulomb gas of the monopoles:
\beqn
S^\mon[\rho] = \frac{g^2_\mon}{2} \sum_{\stackrel{{a,b=1}}{a \neq b}}^N q_a q_b \, D({\bs x}_a-{\bs y}_b) + N S_0\,,
\label{eq:S:Coulomb:gas}
\eeqn
where 
\beqn
S_0 = \frac{g^2_\mon}{2} D({\bs 0})\,,
\label{eq:S:0}
\eeqn
is a divergent term which will be renormalized below. According to Eq.~\eq{eq:D} the Coulomb term in the action~\eq{eq:S:Coulomb:gas} describes long-ranged interaction between the monopole instantons.

A variation of the action~\eq{eq:S:A:rho} with respect to the photon field $A_\mu$ provides us with the usual Maxwell equations for photons: 
\beqn
\partial_\mu F^{\mu\nu}_\ph = 0\,.
\label{eq:dF:2}
\eeqn

The physical content of the model is given by the photons and the monopoles. The model does not possesses any matter fields that bear electric charges. Indeed, the electric and magnetic charges are defined via the Maxwell equation and its dual, respectively:
\beqn
\partial_\mu F^{\mu\nu} & = & j^\nu_{\mathrm{ch}}\,, 
\label{eq:Maxwell:real}\\
\partial_\mu {\widetilde F}^{\mu} & = & j_\mon\,,
\label{eq:Maxwell:dual}
\eeqn
where $j^\mu_{\mathrm{ch}}$ is the electric (charged) current, $j_\mon$ is the magnetic charge density and 
\beqn
{\widetilde F}^{\mu} = \frac{1}{2} \epsilon^{\mu\alpha\beta} F_{\alpha\beta}\,,
\eeqn
is the dual field strength tensor (in 3 spacetime dimensions this tensor is, in fact, a pseudovector).

Substituting~\eq{eq:F:munu} with \eq{eq:F:ph} and \eq{eq:F:mon} into the Maxwell equations~\eq{eq:Maxwell:real} and \eq{eq:Maxwell:dual}, one gets:
\begin{subequations}
\beqn
j^\mu_{\mathrm{ch}}(\bx) & = & 0\,, \\
j_\mon(\bx) & = & g_\mon \rho(\bx)\,.
\eeqn
\label{eq:J:ch:mon}
\end{subequations}
We have used the equation of motion~\eq{eq:dF:1} along with identity~\eq{eq:dF:2} in order to get the first relation in Eq.~\eq{eq:J:ch:mon}. The second relation in \eq{eq:J:ch:mon} comes from the identity $\partial_\mu {\widetilde F}^{\mu}_\ph \equiv 0$ and the explicit form of the monopole field strength tensor~\eq{eq:F:mon}. Thus, we conclude that the spectrum of the model consists of vector photons and instanton-like monopoles only.

The presence of the monopoles is related to the compactness of the gauge group. This property will be evident in the lattice formulation of the theory which will be considered in Sect.~\ref{sec:compact:lattice:QED}.

\subsubsection{Decoupling of photons and monopoles}

The photons and monopoles are the only dynamical degrees of freedom in the model. The partition function
\beqn
Z = \int \cD A  \hskip 3mm \sumint_{\mon}  e^{ - S[A,\rho]}
\label{eq:Z:0}
\eeqn
involves the integration over the photon configurations $A$ and the sum over all monopole configurations encoded via the monopole densities $\rho$. The integration measure over the monopole configurations can be written as follows:
\beqn
\sumint_{\mon} = \sum_{N=0}^\infty \frac{1}{N!} \prod_{a=1}^N \left(\sum_{q_a = \pm 1} \zeta \int d^3 x_a \right).
\label{eq:sumint}
\eeqn
Here the sum goes over the total number of monopoles $N$ (the term with $N=0$ corresponds to a unity in the above sum). The parameter $\zeta$ is the so-called ``fugacity'' which controls the monopole density. We consider the dilute monopole gas so that the monopoles possess a unit charge in elementary magnetic charge~\eq{eq:g:m}, $q_a = \pm 1$. The overlaps between the monopoles are rare and therefore are neglected here.  Basically, in Eq.~\eq{eq:sumint} we integrate over positions $\bx_a$ of all $N$ monopoles, sum over all their magnetic charges $q_a$ and then sum the total monopole number $N$ taking into account the combinatoric degeneracy factor $1/N!$.

Due to the decoupling of the photon and monopole parts of the action~\eq{eq:S:A:rho}, the photon and monopole parts of the partition function~\eq{eq:Z:0} are also decoupled:
\beqn
Z = Z_\ph \cdot Z_\mon\,, 
\label{eq:Z}
\eeqn
where the photon and monopole actions
\beqn
Z_\ph = \int \cD A \, e^{ - S_\ph[A]}\,, 
\label{eq:Z:ph}\\
Z_\mon = \hskip 3mm \sumint_{\mon}  e^{ - S_\mon[\rho]}\,,
\label{eq:Z:mon}
\eeqn
are given in Eqs.~\eq{eq:S:A} and \eq{eq:S:rho}, respectively.

\subsection{Dual formulation of monopole dynamics}

\subsubsection{Coulomb gas as a sign-Gordon model}

The partition function of the photon part has the perturbative Gaussian form~\eq{eq:Z:ph}. The monopole partition function~\eq{eq:Z:mon}, which describes nonperturbative effects, can be reformulated in terms of a nonlinear sine-Gordon model. Following Ref.~\cite{Polyakov:1976fu}
 we rewrite the monopole partition function~\eq{eq:Z:mon} as follows:
\begin{widetext}
\begin{subequations}
\beqn
Z_\mon & & = \hskip 3mm \sumint_{\mon}  e^{ - S_\mon[\rho]}
     \ \mathop{\equiv}_{{\color{black}{\mathrm{(a)}}}} \ \sum_{N=0}^\infty \frac{1}{N!} \prod_{a=1}^N \left(\sum_{q_a = \pm 1} \zeta \int d^3 x_a \right)  \exp\left[- \frac{g^2_\mon}{2} \int d^3 x \, \int d^3 y \, \rho(x) D(x-y) \rho(y)\right] 
\label{eq:Z:mon:derivation:a}\\
& & \mathop{\equiv}_{{\color{black}{\mathrm{(b)}}}} \  C  \int \cD \chi \sum_{N=0}^\infty \frac{1}{N!} \prod_{a=1}^N \left(\sum_{q_a = \pm 1} \zeta \int d^3 x_a \right)  \exp\left\{ - \int d^3 x \, \biggl[\frac{1}{2 g^2_\mon} \bigl(\partial_\mu \chi(x) \bigr)^2 + i \chi(x) \rho (x) \biggr]\right\} \\
& & \mathop{\equiv}_{{\color{black}{\mathrm{(c)}}}} \  C  \int \cD \chi \exp\left\{ - \frac{1}{2 g^2_\mon} \int d^3 x \, \bigl(\partial_\mu \chi(x) \bigr)^2\right\}
\sum_{N=0}^\infty \frac{1}{N!} \prod_{a=1}^N \left(\sum_{q_a = \pm 1} \zeta \int d^3 x_a \, e^{ - i q_a \chi(x_a) }  \right)  \\
& & \mathop{\equiv}_{{\color{black}{\mathrm{(d)}}}} \  C  \int \cD \chi \exp\left\{ - \frac{1}{2 g^2_\mon} \int d^3 x \, \bigl(\partial_\mu \chi\bigr)^2\right\}
\sum_{N=0}^\infty \frac{1}{N!} \left(2 \zeta \int d^3 x \, \cos \chi(x) \right)^N  \\
& & \mathop{\equiv}_{{\color{black}{\mathrm{(e)}}}} \  C  \int \cD \chi \exp\left\{ - \int d^3 x \, \left[\frac{1}{2 g^2_\mon} \bigl(\partial_\mu \chi \bigr)^2 
- 2 \zeta \cos \chi \right] \right\} \equiv C \int \cD \chi \exp\left\{ - \int d^3 x \, \cL_{s}(\chi) \right\} \,,
\label{eq:Z:mon:derivation:e}
\eeqn
\label{eq:Z:mon:derivation}
\end{subequations}
\end{widetext}
where
\beqn
\cL_s = \frac{1}{2 g^2_\mon} \bigl(\partial_\mu \chi \bigr)^2 - 2 \zeta \cos \chi\,,
\label{eq:cL:sine}
\eeqn
is the Lagrangian of the sine-Gordon model. Here and below $C$ stands for an inessential constant parameter.

The steps (a)--(e), which are marked in the chain of relations~\eq{eq:Z:mon:derivation} are as follows: (a) the expression under the exponential of the monopole partition function~\eq{eq:Z:mon}, \eq{eq:S:rho} and \eq{eq:sumint} can be linearized with the help of a real-valued scalar field $\chi$ introduced by the Gaussian integration~(b). Notice that the divergent monopole (self) action~\eq{eq:S:0} can be absorbed (renormalized) into the definition of the fugacity,  $\zeta \to \zeta e^{S_0}$. Then the explicit expression for the monopole density~\eq{eq:rho} can be used (c) to perform the sum (d) over the monopole charges $q_a = \pm 1$. Finally, the the sum over the total monopole number $N$ is converted into the exponent in the last step (e).

Thus, we have represented the theory of monopoles in terms of the field theory with the Lagrangian~\eq{eq:cL:sine}. The latter corresponds to a dual formulation of the Coulomb gas of the monopoles~\eq{eq:S:rho}, in which the dynamics of the original monopole density~\eq{eq:rho} is described by a scalar real-valued field $\chi$.

\subsubsection{Monopole density, photon mass and confinement}

The dual theory~\eq{eq:cL:sine} is very convenient in investigation of nonperturbative properties of associated with the monopole dynamics. First of all, we notice that it is very easy to calculate the mean monopole density $\varrho_\mon \equiv \avr{|\rho|}$ in the Coulomb monopole gas~\eq{eq:Z:0}. We notice that 
\beqn
\varrho_\mon \equiv \avr{N} = \frac{\partial \ln Z_\mon}{\partial \ln \zeta}\,,
\label{eq:rho:mon}
\eeqn
where we used the explicit form of the monopole partition function in Eq.~\eq{eq:Z:mon:derivation:a}. Repeating all steps down to Eq.~\eq{eq:Z:mon:derivation:e} and using the explicit form of the dual Lagrangian~\eq{eq:cL:sine} we get for the monopole density~\eq{eq:rho:mon} the following expression $\varrho_\mon = 2 \zeta \avr{\cos \chi}$. In the leading order the mean monopole density is,
\beqn
\varrho_\mon = 2 \zeta\,,
\label{eq:density}
\eeqn
where we have neglected the quantum corrections due to fluctuations of the dual field $\chi$. 

The field $\chi$ in Eq.~\eq{eq:cL:sine} has the mass
\beqn
m_\ph = g_\mon \sqrt{2 \zeta} \equiv \frac{2\pi \sqrt{2 \zeta}}{g}\,,
\label{eq:m:ph}
\eeqn
where we have used Eq.~\eq{eq:g:m} and expanded the sine-Gordon Lagrangian~\eq{eq:cL:sine} over small fluctuations of the dual field~$\chi$:
\beqn
\cL_s = \frac{1}{2 g^2_\mon} \left[\bigl(\partial_\mu \chi \bigr)^2 + m^2_\ph \, \chi^2 \right] + O(\chi^4) \,.
\label{eq:cL:sine:exp}
\eeqn
Despite the field $\chi$ is associated with the monopoles, the mass~\eq{eq:cL:sine:exp} eventually becomes the mass of the photon (gauge) field. The gauge field $A_\mu$ can be considered to be composed of the regular photon field and singular monopole field which leads to the decomposition of the field strength tensor~\eq{eq:F:munu}. One can show that in the photon-mediated interactions the massless pole of the regular part cancels out and the massive pole of the field $\chi$ determines the interaction range. 

Notice that according to Eq.~\eq{eq:density} the photon mass~\eq{eq:m:ph} can be directly expressed, in the leading order, via the monopole density $\varrho_\mon$ and the electric charge $g$:
\beqn
m_\ph =  \frac{2\pi \sqrt{\varrho_\mon}}{g}\,.
\label{eq:m:ph:rho}
\eeqn
Thus, the monopole gas leads to the nonperturbative mass gap generation in the system. The emergent mass of the photon field is proportional to the square root of the monopole density~\eq{eq:m:ph:rho}.

Equations~\eq{eq:density} and \eq{eq:m:ph:rho} and subsequent relations are valid provided fluctuations of the sine-Gordon field $\chi$ are small, $\avr{\chi^2} \ll 1$, and the relevance of higher-than-quadratic terms in the sine-Gordon model~\eq{eq:cL:sine} is negligible. In terms of monopoles themselves, this condition is realized provided the fluctuations of individual monopoles can be neglected. Since a monopole may affect another monopole at a typical distance of the Debye length $\lambda_D = m_\ph^{-1}$, the fluctuations of individual monopoles are negligible provided the number of monopoles in a unit Debye volume is sufficiently high, $\varrho_\mon \lambda_D^3 \gg 1$. Using Eq.~\eq{eq:m:ph:rho} the applicability requirement may be reformulated as follows: 
\beqn
\varrho^{1/2}_\mon \ll \frac{g^3}{(2\pi)^3}, \qquad\ \mbox{or} \qquad\ \varrho^{1/2}_\mon \; g_\mon^3 \ll 1\,.
\label{eq:applicability}
\eeqn
In other words, our estimations are valid provided the monopole density, expressed in units of magnetic charge~\eq{eq:g:m}, is very small~\eq{eq:applicability}. Therefore Eq.~\eq{eq:applicability} is often called the dilute gas approximation.

Another important property of the compact electrodynamics is the linear confinement of electric charges. It turns out that a pair of static, electrically charged particle and antiparticle separated by sufficiently large distance $R \gg \lambda_D$ experiences the confining potential $V(R) = \sigma R$ which grows linearly with the distance $R$. 
The string tension is given by the following formula:
\beqn
\sigma = \frac{8 \sqrt{2 \zeta}}{g_\mon} \equiv \frac{4 g \sqrt{\varrho_{\mon}}}{\pi}
\label{eq:string:tension}
\eeqn
The compact electrodynamics is one of a few field-theoretical models where the linear confinement property may be proved analytically. Since the confinement property is not a central topic of our study, we leave out the derivation of Eq.~\eq{eq:string:tension} and refer an interested reader to Ref.~\cite{Polyakov:1976fu} for further details.

\section{Casimir effect and monopoles: analytical arguments}
\label{sec:compact:QED:monopoles}

In this section we calculate analytically the effect of dynamical monopoles on the Casimir interaction between two perfectly conducting wires in the limit when the monopole gas is sufficiently dilute~\eq{eq:applicability}. In our derivation we follow the general line of Ref.~\cite{Chernodub:2016owp}. Our analytical calculations of this section will be supplemented by the results of numerical simulations described in the next section.

\subsection{Casimir boundary conditions in integral form}

In two spatial dimensions the basic Casimir problem is formulated for one-dimensional objects. We call these objects as ``wires''. These wires are similar to canonical plates used in the studies of the Casimir (zero-point) interactions in three spatial dimensions.

The effect of a perfectly conducting metallic wire on electromagnetic field is simple. A static and infinitely thin wire makes the tangential component of the electric field vanishing at every point $\bx$ of wire:
\beqn
E_\| (\bx) 
= 0\,.
\label{eq:E:parallel}
\label{eq:F01:3d}
\eeqn
In order to generalize the Casimir boundary condition~\eq{eq:E:parallel} to the most general case of non-static (moving) wires of an arbitrary shape, let us  describe the two-dimensional world surface $S$ of the wire by a vector ${\bar \bx} = {\bar \bx}(\tau,\xi)$ parameterized by the timelike ($\tau$) and spacelike ($\xi$) parameters. The surface element of $S$ can be  described by the singular asymmetric tensor function
\beqn
s_{\mu\nu}(\bx) = \int d \tau \int d \xi \frac{\partial {\bar x}_{[\mu,}}{\partial \tau} \frac{\partial {\bar x}_{\nu]}}{\partial \xi} 
\delta^{(3)}\left(\bx - {\bar \bx}(\tau,\xi)\right), \quad
\label{eq:s:munu:gen}
\eeqn
where $a_{[\mu,} b_{\nu]} = a_\mu b_\nu - a_\nu b_\mu$. Consequently, the boundary condition~\eq{eq:E:parallel} can be rewritten, for any point $\bx$, in the explicitly covariant form:
\beqn
F^{\mu\nu}(\bx) s_{\mu\nu}(\bx) = 0\,,
\label{eq:F:0:cov}
\eeqn
where $F_{\mu\nu}$ is the field strength tensor~\eq{eq:F:munu}.

\begin{figure}[!thb]
\begin{center}
\vskip 3mm
\includegraphics[scale=0.375,clip=true]{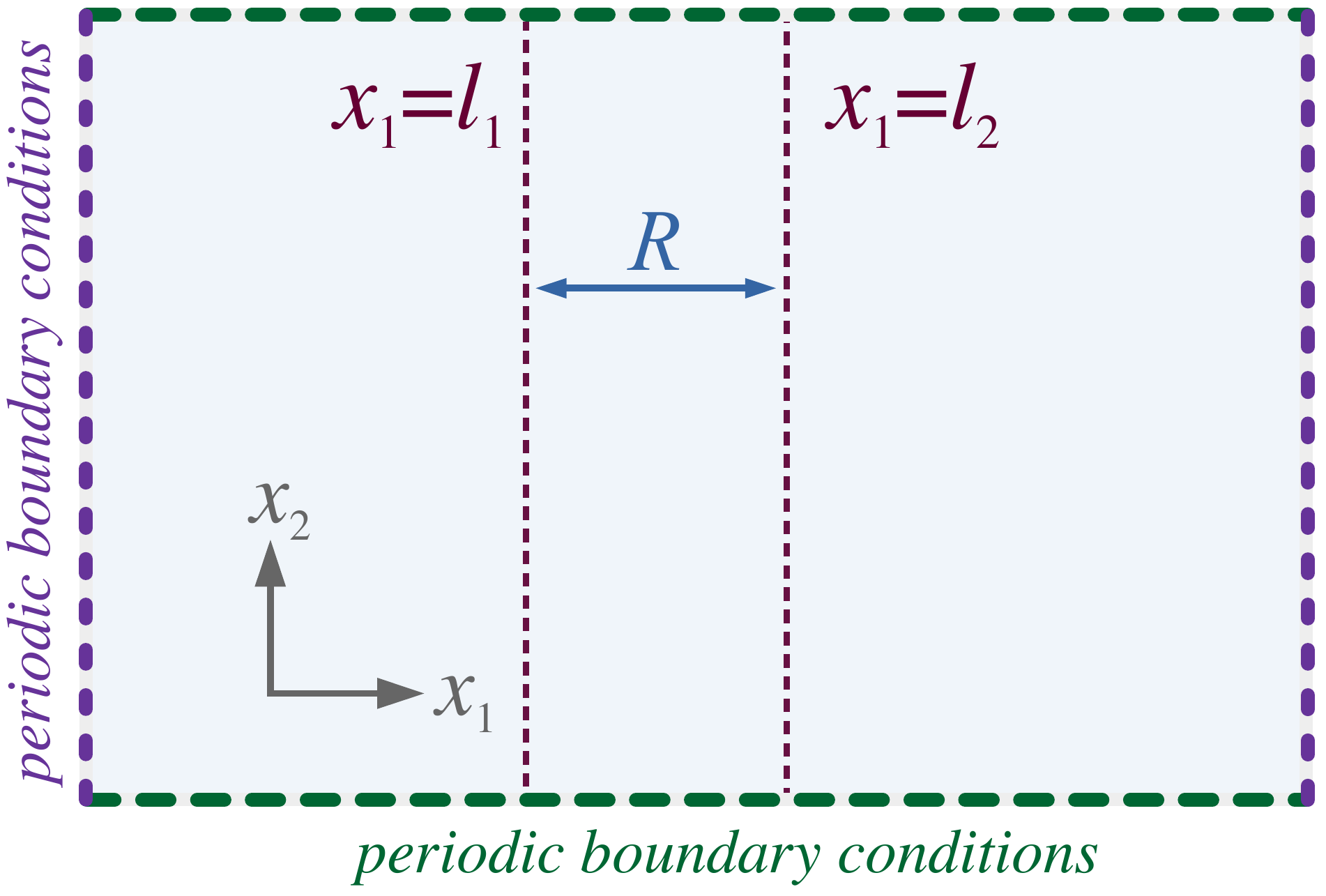}
\end{center}
\vskip -2mm 
\caption{The Casimir problem in two spatial dimensions: the wires $l_1$ and $l_2$ are separated by the distance $R$. The two-dimensional space is compactified into a torus due to periodic boundary conditions.}
\label{fig:geometry:plane}
\end{figure}

In this paper we are interested in zero-point interactions between two parallel wires.  The geometry of our problem is illustrated in Fig.~\ref{fig:geometry:plane}.
In our calculation below the wires are assumed to be perfectly conducting so that they require the tangential component $E_\parallel(x)$ of the electric field to vanish at every point of the wire~$x$. 

A pair of two static wires placed at $x_1 = \pm R/2$ and parallel to the $x_2$ axis can be parametrized with a help of the pair of vector:
\beqn
{\bar \bx}_\pm(\tau,\xi) \equiv (x_1, x_2, x_3) = \left(\pm \frac{R}{2}, \xi,\tau\right)\,,
\label{eq:bx:pm}
\eeqn
where the subscript ``$\pm$'' corresponds to the right/left wire, respectively. The wires are static with respect to the time direction $x_3$. The tensors~\eq{eq:s:munu:gen} for their world surfaces are as follows:
\beqn
s^\pm_{\mu\nu}(\bx) = \left(\delta_{\mu,2} \delta_{\nu,3} - \delta_{\nu,3} \delta_{\mu,2} \right) \delta(x_1 \mp R/2)\,.
\label{eq:s:pm}
\eeqn
The boundary condition for our static straight wires can be read off from Eqs.~\eq{eq:s:pm} and \eq{eq:F:0:cov}:
\beqn
F_{23}(\pm R/2, x_2, x_3) = 0\,.
\label{eq:F23:0}
\eeqn
This relation coincides with Eq.~\eq{eq:E:parallel} because $F_{23} \equiv E_2$ is the component of the electric field that is parallel to the wires.

In the path-integral formalism the Casimir condition~\eq{eq:F:0:cov} may conveniently be implemented with the help of the following $\delta$ functional:
\beqn
\delta_S[F] = \prod_{\bx} \delta\Bigl(F^{\mu\nu}(\bx) s_{\mu\nu}(\bx)\Bigr)\,.
\eeqn
The infinite product of the $\delta$ functions may be rewritten with the help of the functional integration over the Lagrange multiplier $ 
\lambda({\bs x})$:
\beqn
\delta_S[F] & = & \int \cD \lambda  \exp\left[ \frac{i}{2} \int d^3 x \, \lambda(\bx) F^{\mu\nu}(\bx) s_{\mu\nu}(\bx) \right]
\nonumber \\
& \equiv & \int \cD \lambda  \exp\left[ \frac{i}{2} \int d^3 x \, F^{\mu\nu}(\bx) J_{\mu\nu}(\bx;\lambda) \right],
\label{eq:S:F}
\eeqn
where the ``surface tensor''
\beqn
J_{\mu\nu}(\bx;\lambda) =  \lambda(\bx) s_{\mu\nu}(\bx),
\label{eq:J:munu}
\eeqn
is the product of the Lagrange multiplier and the surface tensor~\eq{eq:s:munu:gen}. In our case of the two parallel Casimir plates~\eq{eq:bx:pm} we have:
\beqn
\delta_S[F] = \int \cD \lambda_+ \int \cD \lambda_-  \exp\biggl[ i \int d x_2 \int d x_3 \, \nonumber\\ 
\sum_{a=\pm 1} \lambda_a(x_2,x_3)  F_{23}\left(a \frac{R}{2},x_2,x_3\right) \biggr].
\label{eq:delta:S:plates}
\eeqn
The integration under the exponent is taking place along the two-dimensional world surface, and the integrations over the Lagrange multipliers $\lambda_+$ and $\lambda_-$ enforce the Casimir conditions~\eq{eq:F23:0} at the flat world surfaces for the right ($x_1 = + R/2$) and left ($x_1 = - R/2$) wire, respectively. 

The partition function~\eq{eq:Z}, \eq{eq:Z:ph} and \eq{eq:Z:mon} in the presence of the Casimir surface $S$ is then given by following compact formula:
\beqn
Z_S = \int \cD A \hskip 3mm \sumint_{\mon}  \, e^{ - S_\ph[A] - S_\mon[\rho]} \, \delta_S[F]\,.
\label{eq:Z:S}
\eeqn

\subsection{Casimir boundaries in dual sine-Gordon theory}

Since the field strength tensor contains both the photon and monopole parts, $F = F_\ph + F_\mon$, the decoupling of the photon and monopole partition functions~\eq{eq:Z} is no longer possible in the presence of the Casimir surfaces~\eq{eq:Z:S}. It is clearly seen from the partition function~\eq{eq:Z:S} which can be represented in the following functional form
\beqn
Z_S & = & \int \cD \lambda\, Z_\ph[\lambda] Z_\mon[\lambda], \quad
\label{eq:Z:S:lambda}\\
Z_\ph[\lambda] & = & \int \cD A e^{ - S_\ph[A] + \frac{i}{2} \int d^3 x \, F^{\mu\nu}_\ph(\bx) J_{\mu\nu}(\bx;\lambda)}, \quad 
\label{eq:Z:ph:lambda}\\
Z_\mon[\lambda] & = & \hskip 3mm \sumint_{\mon}  \, e^{- S_\mon[\rho] + \frac{i}{2} \int d^3 x \, F^{\mu\nu}_\mon(\bx) J_{\mu\nu}(\bx;\lambda)}, \quad
\label{eq:Z:mon:lambda}
\eeqn
where the surface current $J_{\mu\nu}$ is given in Eq.~\eq{eq:J:munu}.

Performing the Gaussian integration over the photon field $A_\mu$, we get for the photon part~\eq{eq:Z:ph:lambda} the following expression~\cite{Chernodub:2016owp}:
\beqn
& & Z_\ph[\lambda] {=} \int \cD A \, \exp\left[ \int d^3 x \, \left(- \frac{1}{4} F^2_{\mu\nu} + i A_\mu J^\mu\right)\right] \nonumber \\
& & {=} \, C \exp\left[ - \frac{1}{2} \int d^3 x \, d^3 y \, J_\mu(\bx; \lambda) D(\bx - \by) J_\mu(\by; \lambda) \right]\!, \qquad 
\label{eq:Z:ph:lambda:1}
\eeqn
where $C$ stands for an inessential constant which will be omitted below. 

The surface tensor~\eq{eq:J:munu} enters Eq.~\eq{eq:Z:ph:lambda:1} via the conserved vector 
\beqn
J_\mu(\bx; \lambda) = \partial^\nu J_{\mu\nu}(\bx;\lambda)\,, \qquad \partial^\mu J_\mu(\bx;\lambda) = 0\,.
\label{eq:J:mu}
\eeqn
Using Eqs.~\eq{eq:s:pm} and \eq{eq:J:munu} we find that for the two parallel straight wires~\eq{eq:bx:pm} the current~\eq{eq:J:mu} is given by the following expression:
\beqn
J_\mu = \sum_{a = \pm } \delta\biggl(x_1 - \frac{a R}{2} \biggr) \biggl(\delta_{\mu2} \frac{\partial \lambda_a}{\partial x_3} - \delta_{\mu3}  \frac{\partial \lambda_a}{\partial x_2} \biggr)\,,
\label{eq:J:mu:flat}
\eeqn
where $\lambda_\pm = \lambda_\pm (x_2,x_3)$ are the Lagrange multipliers associated with the left and right plates, respectively. 

The monopole part~\eq{eq:Z:mon:lambda} can be evaluated in analogy with the chain of transformations~\eq{eq:Z:mon:derivation}. Using the explicit form of the monopole field strength tensor~\eq{eq:F:mon} we get the following representation of the monopole partition function in the sine-Gordon form:
\beqn
Z_\mon[\lambda] = \int \cD \chi \exp\left\{ - \int d^3 x \, \cL_{s}(\chi;\lambda) \right\} \,,
\label{eq:Z:mon:chi}
\eeqn
where the kinetic term in the sine-Gordon action
\beqn
\cL_s(\chi;\lambda) = \frac{1}{2 g^2_\mon} \Bigl[\partial_\mu \bigl(\chi(\bx)  - q(\bx,\lambda)\bigr)\Bigr]^2 {-} 2 \zeta \cos \chi (\bx), \qquad
\label{eq:cL:lambda:sine}
\eeqn
is modified by a ``curl'' of the surface tensor~\eq{eq:J:munu}:
\beqn
q(\bx,\lambda) = \frac{g_\mon}{2} \int d^3 y \, D(\bx - \by) \epsilon_{\alpha\mu\nu} \partial^\alpha J^{\mu\nu} (\by;\lambda)\,.
\label{eq:q:general}
\eeqn

For two parallel straight lines~\eq{eq:bx:pm} the explicit form of the scalar function~\eq{eq:q:general} is as follows:
\beqn
q(\bx) & = & g_\mon \int d y_2 d y_3 \sum_{a=\pm} \lambda_a(y_2,y_3) \\
& & \cdot \frac{\partial}{\partial x_1} D\left(x_1 - \frac{a R}{2},x_2 - y_2,x_3 - y_3\right). \nonumber
\label{eq:q:flat}
\eeqn

Finally, we substitute the photon~\eq{eq:Z:ph:lambda:1} and monopole~\eq{eq:Z:mon:chi} partition functions into Eq.~~\eq{eq:Z:S:lambda} and get for the total partition function in the presence of the Casimir boundaries:
\begin{widetext}
\beqn
Z_S {=} \int \cD \lambda \cD \chi  
\exp\left[ - \frac{1}{2} \int d^3 x \, d^3 y \, J_\mu(\bx; \lambda) D(\bx - \by) J_\mu(\by; \lambda) 
{-} \int d^3 x \, \left(\frac{1}{2 g^2_\mon} \Bigl[\partial_\mu \bigl(\chi(\bx)  - q(\bx,\lambda)\bigr)\Bigr]^2 {-} 2 \zeta \cos \chi (\bx)\right)\right]\!,
\qquad 
\label{eq:Z:lambda:tot}
\eeqn
\end{widetext}
where the current~$J_\mu$ and the scalar $q$ are given in Eqs.~\eq{eq:J:mu} and \eq{eq:q:flat}, respectively. 

It is interesting to notice the effect of the Casimir boundary condition on the dual sine-Gordon field $\chi$ is quite nontrivial. One could naively expect that the requirement of vanishing of the electric field~\eq{eq:E:parallel} at the surface of the wire would lead either to Dirichlet or to Neumann boundary condition for the dual field $\chi$ at the worldsheet of the wire (the field $\chi$ or its normal derivative, respectively, would vanish at the position of the wire). According to our result~\eq{eq:Z:lambda:tot} this naive expectation is not true.

In the partition function~\eq{eq:Z:lambda:tot} the nonlinear term in sine-Gordon field may be expanded in powers of the scalar field $\chi(\bx)$ while the local interaction terms $\chi^{n}(\bx)$ with $n \geqslant 4$ may be neglected in the leading order of the dilute gas approximation~\eq{eq:applicability}. Therefore the functional in the exponential becomes quadratic both in the field $\chi$ and in the field $\lambda$ so that the corresponding integrals become Gaussian and thus can be easily evaluated. Below we calculate them explicitly for the case of two parallel static wires.

\subsection{Casimir potential for parallel wires}

For two static straight wires separated by the distance $R$ the density of the Casimir energy $V(R)$ per unit length of the wire is given by the following formula
\beqn
V(R) = - \frac{1}{\cA} \ln Z_{W_R}\,,
\label{eq:Z:W:R}
\eeqn
where $\cA = T L$ is the area of the worldsheet of each of the wires. Both the length of the wire $L$ and the time of their existence $T$ are assumed to be very (infinitely) long. The partition function $Z_{W_R}$ is explicitly given by Eq.~\eq{eq:Z:lambda:tot} where the surface $S = W_R \equiv W_{R_+} \cup W_{R_-}$ is represented by two flat sheets corresponding to the parallel wires. For the sake of convenience, below we repeat a known derivation of the Casimir potential~\eq{eq:Z:W:R} in the absence of the monopoles and then we evaluate the effect of the monopole gas. 

\subsubsection{Casimir energy in the absence of monopoles}

The density of monopoles~\eq{eq:density} is proportional to the fugacity parameter $\zeta$. By setting $\zeta=0$ we remove the monopoles from the ensembles. As a consequence, the nonlinear term in the sine-Gordon Lagrangian~\eq{eq:cL:sine} disappears, and
the partition function~\eq{eq:Z:lambda:tot} becomes independent of the functional $q$ because it can now be absorbed in the sine-Gordon field $\chi$ by the shift $\chi \to \chi + q$. Then in Eq.~\eq{eq:Z:lambda:tot} the field $\chi$ can be integrated out exactly:
\beqn
Z_S = \int \cD \lambda \, e^{ - \frac{1}{2} \int d^3 x \, d^3 y \, J_\mu(\bx; \lambda) D(\bx - \by) J_\mu(\by; \lambda)}\,. 
\label{eq:Z:lambda:nomon}
\eeqn
Next, using the explicit form~\eq{eq:J:mu:flat} of the current $J_\mu(\bx; \lambda)$ one gets for the partition function~\eq{eq:Z:lambda:nomon} of the plates:
\beqn
Z_{W_R} = \int \cD \Lambda \, e^{ - \frac{1}{2} \int d^2 x \, d^2 y \, \Lambda^T(\vx) {\widehat K}(\vx - \vy) \Lambda(\vy)}\,,
\label{eq:Z:lambda:nomon:1}
\eeqn
where we introduced the two-dimensional vector on the worldsheets of the wires, $\vx = (x_2,x_3)$. We also introduced the vector field:
\beqn
\Lambda(\vx) = 
\left(
\begin{array}{c}
\lambda_+ (\vx) \\
\lambda_- (\vx)
\end{array}
\right)\,,
\eeqn
and the matrix
\beqn
{\widehat K}(\vx) & = & \left(\frac{\partial^2}{\partial x_2^2} + \frac{\partial^2}{\partial x_3^2}\right)
\label{eq:K} \\
& & \left(
\begin{array}{ll}
D(0, x_2, x_3) & D(- R, x_2, x_3) \\
D(+ R, x_2, x_3) & D(0, x_2, x_3) 
\end{array}
\right)\,, \nonumber
\eeqn
where the function $D(\bx)$ given in Eq.~\eq{eq:D}.

The Gaussian integral~\eq{eq:Z:lambda:nomon:1} is given (up to a multiplicative constant) by the determinant of the operator~\eq{eq:K}:
\beqn
Z_{W_R} = {\mathrm{det}}^{-1/2}\, {\widehat K}\,.
\label{eq:Z:lambda:nomon:2}
\eeqn
The density of the Casimir energy is given by Eqs.~\eq{eq:Z:W:R}:
\beqn
V(R) = \frac{1}{2 \cA} {\mathrm{Tr}} \log {\widehat K}.
\label{eq:A:V:R}
\eeqn

Formally, Eq.~\eq{eq:A:V:R} can be written as a sum over all eigenvalues~$\kappa_i$
\beqn
{\mathrm{Tr}} \log {\widehat K} \equiv \sum_i \log \kappa_i\,,
\label{eq:Tr:log:formal}
\eeqn
of the integral operator ${\widehat K}$:
\beqn
{\widehat K} L_i = \kappa_i L_i\,,
\label{eq:eigen:K}
\eeqn
which can be rewritten in the explicit form as follows:
\beqn
\int d^2 y \, {\widehat K}(\vx - \vy) L(\vy) = \kappa L(\vx)\,.
\label{eq:K:L:xy}
\eeqn

It is convenient to represent the operator $\widehat K$, given by Eqs.~\eq{eq:K} and \eq{eq:D}, and its eigenvalues $L_i(\vx)$ as the Fourier integrals,
\beqn
{\widehat K}(\vx) & = &- \int \frac{d^3 k}{(2\pi)^3} \frac{p_2^2 + p_3^2}{p_1^2 + p_2^2 + p_3^2}
\left(
\begin{array}{ll}
1 & e^{-i p_1 R} \\
e^{i p_1 R} & 1 
\end{array}
\right),\qquad 
\label{eq:K:p} \\
L_i(\vx) & = & \int \frac{d^2 q}{(2 \pi)^2} L_i(\vq) e^{i \vq \vx}\,.
\label{eq:L:x}
\eeqn
Then we substitute Eqs.~\eq{eq:K:p} and \eq{eq:L:x} into Eq.~\eq{eq:K:L:xy}, integrate over $\vy$ and get the following eigenvalue equation:
\beqn
\int \frac{d^2 q}{(2 \pi)^2} \left[{\widehat Q}(R,{\vec q}) - \kappa_i \right] L_i(\vq) e^{i \vq \vx} = 0\,,
\label{eq:eigen:1}
\eeqn
where
\beqn
{\widehat Q}(R,{\vec q}) = - \frac{|\vec q|}{2}
\left(\begin{array}{cc}
1 & e^{- |{\vec q}| R} \\
e^{- |{\vec q}| R} & 1 
\end{array}\right)\,.
\label{eq:hat:Q}
\eeqn

Since Eq.~\eq{eq:eigen:1} should be valid for all vectors $\vq$, we arrive to the following equation for the eigenmodes:
\beqn
\left[{\widehat Q}(R,{\vec q}) - \kappa_i \right] L_i(\vq) = 0\,,
\label{eq:eigen:2}
\eeqn
which has the following eigenvalues $\kappa_i \equiv \kappa_\pm (\vq)$:
\beqn
\kappa_\pm (\vq) = - \frac{|\vq|}{2} \left(1 \pm e^{- |\vq| R}\right)\,.
\label{eq:kappa:pm}
\eeqn
The solutions are characterized by the discrete index $\pm$ and continuous parameter $\vq$. The phase space associated with these variables is
\beqn
{\mathrm{Tr}}_{\vq} \equiv \cA \int \frac{d^2 q}{(2\pi)^2} \sum_\pm\,,
\label{eq:Tr:q}
\eeqn
where $\cA$ is the area in the transverse $\vx \equiv (x_2,x_3)$ plane.

Substituting Eqs.~\eq{eq:kappa:pm} and \eq{eq:Tr:q} into \eq{eq:A:V:R} and evaluating the integrals explicitly we get the known result for the density of the Casimir energy in the absence of monopoles:
\beqn
V_{\mathrm{Cas}}(R) & = & \frac{1}{2} \int \frac{d^2 q}{(2\pi)^2} \log \left(1 - e^{- 2 |\vq| R}\right) \nonumber \\
& & =  - \frac{\zeta(3)}{16 \pi} \frac{1}{R^2}\,,
\label{eq:V:Cas:R}
\eeqn
where $\zeta(x)$ is the zeta-function with $\zeta(3) \approx 1.20206$. As usual, in our calculation a divergent $R$-independent contribution to the potential $V(R)$ has been been omitted.

\subsubsection{Casimir energy in the presence of monopoles}

Now let us consider the case of the finite monopole density which is given by a nonzero fugacity parameter $\zeta$ in Eq.~\eq{eq:Z:lambda:tot}. 
We expand in Eq.~\eq{eq:Z:lambda:tot} the sine-Gordon action over the small fluctuations of the dual field~\eq{eq:cL:sine:exp}, and arrive to the following representation of the partition function:
\beqn
Z_S & = & \int \cD \lambda \cD \chi \exp\biggl\{ \frac{1}{2} \left(J_\mu(\lambda), \Delta^{-1} J_\mu(\lambda)\right) \nonumber\\
&&  - \frac{1}{2 g^2_\mon} \int d^3 x \, \Bigl[\Bigl(\partial_\mu \bigl(\chi - q(\lambda) \bigr)\Bigr)^2 + m_\ph^2 \chi^2 \Bigr]\biggr\}\,.
\qquad \label{eq:Z:lambda:2}
\eeqn
Here and below we use the following shorthand notations:
\beqn
& & D = - \Delta^{-1}\,, \\
& & \int d^3 y \, D(\bx - \by) A(\by) = - \Delta^{-1} A\,, \\
& & \int d^3 x \, d^3 y \, A(\bx) D(\bx - \by) B(\by) = - \left(B, \Delta^{-1} A  \right).
\eeqn
In Eq.~\eq{eq:Z:lambda:2} we omit the $O(\chi^4)$ interaction terms which, in the dilute gas approximation~\eq{eq:applicability}, represent next to the leading order corrections.

Integrating out the Gaussian field $\chi$ in Eq.~\eq{eq:Z:lambda:2} we get
\beqn
Z_S & = & \int \cD \lambda \cD \chi \exp\biggl\{ \frac{1}{2} \left(J_\mu(\lambda), \Delta^{-1} J_\mu(\lambda)\right) \nonumber\\
&&  - \zeta \biggl(q(\lambda), \frac{\Delta}{-\Delta + m_\ph^2} q(\lambda) \biggr) \Bigr]\biggr\}\,.
\qquad \label{eq:Z:lambda:3}
\eeqn
Then, for the case of the flat world surfaces, we use Eqs.~\eq{eq:J:mu:flat} and \eq{eq:q:flat} and rewrite Eq.~\eq{eq:Z:lambda:3} in the form similar to Eq.~\eq{eq:Z:lambda:nomon:1}:
\beqn
Z_{W_R} = \int \cD \Lambda \, e^{ - \frac{1}{2} \int d^2 x \, d^2 y \, \Lambda^T(\vx) {\widehat K}_{m_{\mathrm{ph}}}(\vx - \vy) \Lambda(\vy)}\,,
\label{eq:Z:lambda:mon:1}
\eeqn
where the operator ${\widehat K}_{m_{\mathrm{ph}}}$ is a massive analogue of the ${\widehat K}$ operator in Eq.~\eq{eq:K}:
\beqn
{\widehat K}_{m_\ph}(\vx) & = & \left(\frac{\partial^2}{\partial x_2^2} + \frac{\partial^2}{\partial x_3^2}\right)
\label{eq:K:m:ph} \\
& & \left(
\begin{array}{ll}
D_{m_\ph}(0, x_2, x_3) & D_{m_\ph}(- R, x_2, x_3) \\
D_{m_\ph}(+ R, x_2, x_3) & D_{m_\ph}(0, x_2, x_3) 
\end{array}
\right). \nonumber
\eeqn
The function ,
\beqn
D_{m_{\ph}}({\bs x}) = \int \frac{d^3 k}{(2 \pi)^3} \frac{e^{i {\bs k} \bx}}{k^2+m_\ph^2}\,.
\label{eq:D:mph}
\eeqn
is a Green function of a massive scalar field:
\beqn
(- \Delta + m_\ph^2) D_{m_\ph}(\bx) = \delta(\bx)\,.
\eeqn
The photon mass $m_\ph$ is given in Eqs.~\eq{eq:m:ph} and \eq{eq:m:ph:rho}.

\begin{figure}[!thb]
\begin{center}
\vskip 5mm 
\includegraphics[scale=0.5,clip=true]{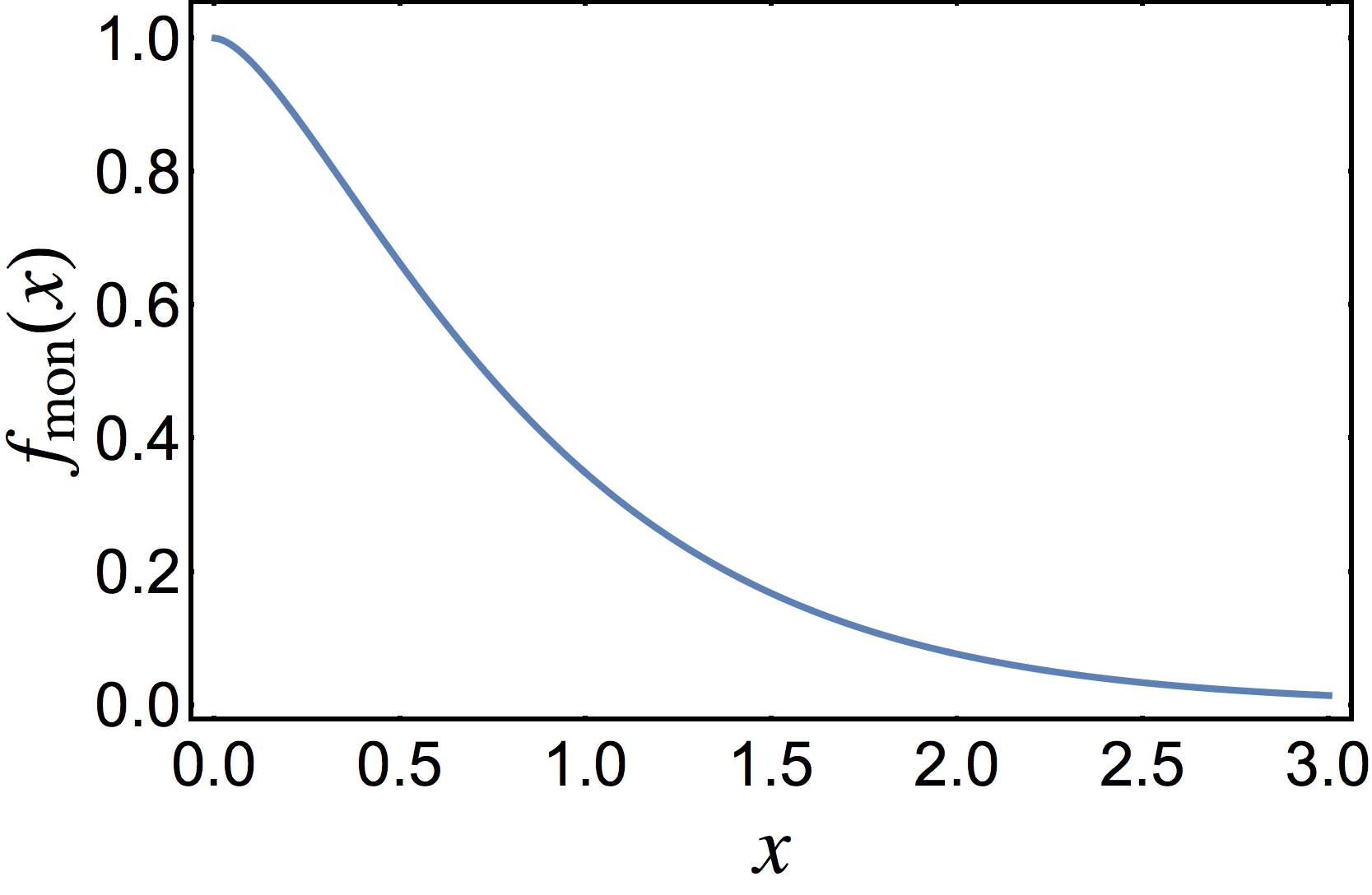}
\end{center}
\vskip -5mm 
\caption{The function $f_\mon$ in Eq.~\eq{eq:f:x}.}
\label{fig:f:x}
\end{figure}

Following all steps made of previous section we arrive to the following expression for the Casimir energy density in the presence of monopoles:
\beqn
V^\mon_{\mathrm{Cas}}(R,m_\ph) & = & \frac{1}{2} \int \frac{d^2 q}{(2\pi)^2} \log \left(1 - e^{- 2 \sqrt{\vq^{\;2} + m_\ph^2} R}\right) \nonumber\\ 
& = &  - \frac{\zeta(3)}{16 \pi} \frac{1}{R^2}  \, f_\mon \left(m_\ph R\right) \,.
\label{eq:V:Cas:R:massive}
\eeqn
where the function
\beqn
f_\mon(x) = - \frac{2 x^2}{\zeta(3)} \int_0^\infty d y \, \log \left( 1 - e^{- 2 x \sqrt{y+1}} \right)\,.
\label{eq:f:x}
\eeqn
is shown in Fig.~\ref{fig:f:x}.

The monopole density $\varrho_\mon$ enters the Casimir energy via the photon mass~\eq{eq:m:ph:rho} in terms of the function 
\beqn
f_\mon\left(m_\ph R\right) \equiv \frac{V_{\mathrm{Cas}}(R,m_\ph)}{V_{\mathrm{Cas}}(R,0)}\,,
\eeqn
where $V_{\mathrm{Cas}}(R) \equiv V_{\mathrm{Cas}}(R,0)$ is the Casimir energy density in the absence of the monopoles~\eq{eq:V:Cas:R}.

At small distances between the wires (or, equivalently, at small monopole density, $R \, m_\ph \ll 1$) the function $f_\mon$ is close to unity, $f_\mon(x) = 1 + O(x^2)$. Therefore in this case the Casimir energy~\eq{eq:V:Cas:R:massive} is close to the Casimir energy in the standard non-compact electrodynamics where the monopoles are absent~\eq{eq:V:Cas:R}. However the function $f_\mon$ decays exponentially  at large value of its argument,
\beqn
f_\mon(x) = \frac{2 x}{\zeta(3)} e^{- 2 x} + \dots \,, \qquad x \gg 1\,,
\label{eq:f:mon:expansion}
\eeqn
indicating that at large monopole densities and/or at large wire separations, $R m_\ph \gg 1$, the Casimir energy density should be exponentially suppressed. 

Thus we come to the conclusion that in the dilute gas approximation the presence of the dynamical monopoles leads to suppression of the Casimir effect at large separation between the wires. This nonperturbative effect effect does not come totally unexpected in view of the mass gap generation in the Coulomb gas of monopoles.

In the next Section we study the Casimir interaction numerically in the lattice formulation of the theory.  The numerical approach allows us to explore certain unexpected monopole effects that are beyond the dilute gas approximation~\eq{eq:applicability}.

\section{Compact lattice electrodynamics}
\label{sec:compact:lattice:QED}

\subsection{Action, photons and monopoles}

The lattice version of the action~\eq{eq:L} of the compact electrodynamics in three space-time dimensions is given by the sum over elementary plaquettes $P $ of the lattice:
\beqn
S[\theta] = \beta \sum_P \left(1 - \cos \theta_P \right)\,,
\label{eq:S}
\eeqn
A plaquette $P \equiv P_{x,\mu\nu}$ is determined by a position~$x$ of one of its corners and its orientation given by two orthogonal vectors $\mu < \nu$ in the plaquette plane with $\mu,\nu = 1, 2, 3$. The plaquette angle 
\beqn
\theta_{P_{x,\mu\nu}} = \theta_{x,\mu} + \theta_{x+\hat\mu,\nu} - \theta_{x+\hat\nu,\mu} - \theta_{x,\nu}\,,
\label{eq:theta:P}
\eeqn
is constructed from elementary angles 
\beqn
\theta_{x,\mu} \in [-\pi,+\pi)\,,
\label{eq:angles:lattice}
\eeqn
which belong to the links of the lattice starting at the point $x$ and pointing towards the direction $\mu$. The lattice coupling constant
\beqn
\beta = \frac{1}{g^2 a}\,,
\label{eq:beta:3D}
\eeqn
is determined by the length of the elementary lattice link (lattice spacing) $a$ and the electric charge $g$. The dimensionality of the continuum coupling $g$ in $3$ dimensional spacetime $[g] = {\text{mass}}^{1/2}$, so that the lattice coupling $\beta$ in Eq.~\eq{eq:beta:3D} is a dimensionless quantity. The relation~\eq{eq:beta:3D} is valid in the regime of the weak coupling $g$ (large $\beta$) where the expansion of the action in terms of small field fluctuations, and consequently the comparison with the continuum action~\eq{eq:L}, is possible.

The angular variable $\theta_{x\mu}$ is a lattice version of the continuum  Abelian gauge field $A_\mu$, $\theta_{x\mu} = a A_{\mu}(x)$. In the continuum limit the lattice spacing approaches zero, $a \to 0$, and the plaquette variable~\eq{eq:theta:P} reduces, for finite values of the gauge field $A_\mu$, to the continuum field strength tensor~\eq{eq:F:munu}: 
\beqn
\theta_{P_{x,\mu\nu}} = a^2 F_{\mu\nu}(x) + O(a^4)\,.
\label{eq:theta:P:continuum}
\eeqn
Substituting this equation into the lattice action~\eq{eq:S}, expanding over the powers of the lattice spacing $a$ and keeping the leading term only, we get the action of the continuum U(1) gauge theory~\eq{eq:S:continuum}.

The partition function of the model,
\beqn
\cZ = \prod_{l} \int_{-\pi}^\pi d \theta_l \, e^{-S[\theta]}\,, 
\label{eq:Z:lat}
\eeqn
includes integration over all angular link variables~\eq{eq:angles:lattice}. The action~\eq{eq:S} is invariant under the $2 \pi$ shifts of the plaquette variable,
\beqn
\theta_P \to \theta_P + 2 \pi n\,, \qquad \quad n \in \Z\,,
\label{eq:theta:P:shifts}
\eeqn
indicating that the lattice field strengths $\theta_P$ and $\theta_P' = \theta_P + 2 \pi$ that are different by a $2 \pi$ shift~\eq{eq:theta:P:shifts} are physically equivalent indicating that the Abelian gauge group is a compact manifold (hence the name, ``compact electrodynamics'' or ``compact gauge theory''). In the continuum limit~\eq{eq:theta:P:continuum} these $2 \pi$ shifts become singular functions proportional to $2 \pi/a^2$ where $a \to 0$ is a vanishing lattice spacing. These shifts corresponds to the Dirac sheets which are world-lines of the Dirac strings attached to the Abelian monopoles. The positions of the Dirac strings are gauge-dependent so that the Dirac strings themselves are not gauge invariant objects. However the ends of the Dirac strings, monopoles, are physical, gauge-invariant topological defects. Thus, the compactness of the model leads to singular configurations which appear as Abelian monopoles and lead to the decomposition of the continuum field-strength tensor to the regular (photon) and singular (monopole) parts~\eq{eq:F:munu}. A comprehensive review on compact (gauge) fields and topological defects can be found in Ref.~\cite{ref:book:Kleinert}.

In the three-dimensional compact electrodynamics the monopoles are pointlike (instanton-like) objects. In the continuum limit their density is given by Eq.~\eq{eq:rho}. On the lattice, the monopoles are living on three-dimensional cubes $C_x$. Their local magnetic charge density
\beqn
\rho_x = \frac{1}{2\pi} \sum_{P \partial C_x} {\bar \theta}_P\,,
\label{eq:rho:lattice}
\eeqn
is nothing but a divergence of the physical part of the lattice field-strength tensor~\eq{eq:theta:P}
\beqn
{\bar \theta}_P = \theta_P + 2 \pi k_P \in [-\pi,\pi), \qquad k_P \in \Z,
\label{eq:bar:theta}
\eeqn
where the integer number $k_P$ is chosen in such a way that the plaquette angle ${\bar \theta}_P$ belongs to the canonical interval. One can show that the lattice monopole density~\eq{eq:rho:lattice} is an integer number, $\rho_x \in \Z$. In fact, Eq.~\eq{eq:rho:lattice} is a divergence of a curl field which is zero for the regular (noncompact) gauge fields and non-zero for singular (compact) fields. The monopoles were studied intensively both in Abelian and non-Abelian lattice gauge theories~\cite{Chernodub:1997ay}.

\subsection{Casimir boundary conditions on the lattice}

The Casimir boundary conditions on the lattice were formulated in various dimensions for Abelian and non-Abelian gauge theories in Ref.~\cite{Chernodub:2016owp}. Here we briefly mention the results relevant to our context. 

In (2+1) dimensions the Casimir boundary conditions force a tangential component of the electric field to vanish at each wire~\eq{eq:F:0:cov}.  In the lattice gauge theory this boundary condition corresponds to the vanishing of the field strength tensor~\eq{eq:theta:P} -- up to the discrete compact transformations~\eq{eq:theta:P:shifts} -- at the set of the plaquettes $P \in \plane$ that belongs to the world-surfaces of the wires.

We consider two static straight wires directed along the $x_2$ axis and separated along the $x_1$ direction, $x_1 = l_1$ and $x_1 = l_2$, Fig.~\ref{fig:geometry:plane}. The $x_3$ axis is associated with the Euclidean ``time'' direction. In the case of an ideal metal, the corresponding boundary condition is given by the lattice version of Eq.~\eq{eq:F01:3d}:
\beqn
\cos\theta_{x,23} = 1\,,
\label{eq:F01:latt:3D}
\eeqn
where $x = (x_1,x_2,x_3)$ with fixed $x_1=l_1,l_2$ and all possible $x_2$ and $x_3$. 

The simplest way to implement the boundary condition~\eq{eq:F01:latt:3D} is to add a set of Lagrange multipliers which will force the plaquettes belonging to the plane $\plane$ to vanish. To this end the standard $U(1)$ action~\eq{eq:S} can be changed as follows:
\beqn
S_{\varepsilon}[\theta;\plane] = \sum_P \beta_P(\varepsilon) \cos \theta_P\,,
\label{eq:S:beta}
\eeqn
where the plaquette-dependent gauge coupling is
\beqn
\beta_{P_{x,\mu\nu}} (\varepsilon) = \beta \bigl[1 + (\varepsilon - 1)\, & & (\delta_{\mu,2} \delta_{\nu,3} + \delta_{\mu,3} \delta_{\nu,2}) \nonumber \\
& & \cdot \left(\delta_{x,l_1} + \delta_{x,l_2}\right)\bigr]\,.
\label{eq:beta:P:3d}
\eeqn
is a function of the dielectric permittivity $\varepsilon$. At $\varepsilon = 1$ the wires are absent while in the limit $\varepsilon \to \infty$ the components of the physical lattice field-strength tensor~\eq{eq:bar:theta} vanish at the word-surfaces of the wires, ${\bar \theta}_{x,23} = 0$ as required by Eq.~\eq{eq:F01:latt:3D}.
The partition function~\eq{eq:Z:lat} of the model in the presence of the Casimir plates becomes as follows:
\beqn
\cZ[\plane] & = & \lim_{\varepsilon \to + \infty} \cZ_\varepsilon[\plane]\,, 
\label{eq:Z:Casimir:1}
\\
\cZ_\varepsilon[\plane] & = & \int \cD \theta \, e^{-S_\varepsilon[\theta;\plane]}\,.
\label{eq:Z:Casimir:2}
\eeqn
In our simulations we realize the case of perfectly conducting wires by considering the limit of large permittivity $\varepsilon \to \infty$ in which a component of the electric field parallel to the wire vanishes~\eq{eq:F01:3d} thus mimicking an ideal metal. In two spatial dimensions the magnetic permeability does not exist and a wire with infinite static dielectric permittivity affects the electromagnetic field in the same way as an ideal metal (cf. Section 5.1 of Ref.~\cite{ref:Bogdag}). 
Our formulation also allows us to consider the system at finite permittivity $\varepsilon$ described by the partition function~\eq{eq:Z:Casimir:2}.

\section{Casimir effect and monopoles: numerical simulation}
\label{sec:cQED:3D}

\subsection{Numerical setup}

Following our previous study~\cite{Chernodub:2016owp} we consider a symmetric $24^3$ cubic lattice which corresponds to a zero-temperature theory. We impose the periodic boundary conditions at the opposite sides of the lattice along all three directions. The two parallel static straight wires are located at the positions $x_1 = l_1$ and $x_1 = l_2$ separated by the distance $R = |l_2 - l_1|$, Fig.~\ref{fig:geometry:plane}. The wires are implemented via the space/orientation-dependent gauge coupling~\eq{eq:beta:P:3d} in the action of the theory~\eq{eq:S:beta}. The wires divide the $x_1$ axis into two, generally inequivalent, intervals, $R$ and $L_s - R$. Due to the periodic boundary conditions all calculated $R$-dependent quantities (potentials, densities, etc) should be invariant under the spatial flip in the $x_1$ direction, $R \to L_s - R$.

\vskip 2mm
  \begin{table}[!ht]
    \centering
    \begin{tabular}{|l|c|}
      \hline Trajectories per one value of $\varepsilon$ & $2 \times 10^5$ \\
      \hline Trajectories for thermalization              & $2 \times 10^4$\\
      \hline Overrelaxation steps between trajectories    & $5$\\
      \hline Lattice size & $24^3$ \\
      \hline Range of gauge coupling & $\beta = 1.0\, \sim\, 1.9$ \\
      \hline Values of permittivity $\varepsilon$ per single value of $\beta$ & $\approx 33$ \\ 
      \hline
    \end{tabular}
    \caption{Basic simulation parameters.}
    \label{tabl:simparam}
  \end{table}
\vskip 2mm

Our numerical tools are the same as in Ref.~\cite{Chernodub:2016owp}. We generate gauge-field configurations using a Hybrid Monte Carlo algorithm which combines the molecular dynamics approach with standard Monte-Carlo methods~\cite{ref:Gattringer}. The molecular-dynamics component utilizes a second-order minimum norm integrator~\cite{ref:Omelyan} with several time scales~\cite{ref:Sexton}. The use of different timescales allows us to equilibrate the integration errors accumulated at and outside worldsheets of the wires at which the Casimir boundary conditions are imposed. Long autocorrelation lengths in Markov chains are eliminated by overrelaxation steps which separate gauge field configurations far from each other~\cite{ref:Gattringer}.  We use a self-tuning adaptive algorithm in order to control the acceptance rate of the Hybrid Monte-Carlo in a reasonable range between 0.70 and 0.85. The parameters of our simulations are presented in Table~\ref{tabl:simparam}. Other details of simulations may be found in Ref.~\cite{Chernodub:2016owp}.

\subsection{Monopoles in the absence of wires}

The monopole density falls off very quickly with increase of the lattice coupling $\beta$ (or, equivalently, with decrease of the coupling $g$ in the continuum limit). In our previous calculations~\cite{Chernodub:2016owp} we were working in the region of sufficiently large lattice gauge coupling $\beta \geqslant 3$ where the density of the monopoles was extremely small. In this region the theory becomes effectively noncompact since strong values of the lattice strength tensor $\theta_P \sim \pi + 2 \pi n$ (with $n \in \Z$) are practically unaccessible at so large coupling $\beta$. Therefore we have found no measurable monopole effects and we have proven reliability of our lattice methods by demonstrating reproducibility of well-known theoretical results. In the present we would like to study the effect of Abelian monopoles on the Casimir forces and therefore we choose the region of small lattice gauge coupling~$\beta$ which corresponds, according to Eq.~\eq{eq:beta:3D}, to the strong coupling region in term of the continuum electric charge~$g$.

\begin{figure}[!thb]
\begin{center}
\vskip 3mm
\includegraphics[scale=0.55,clip=true]{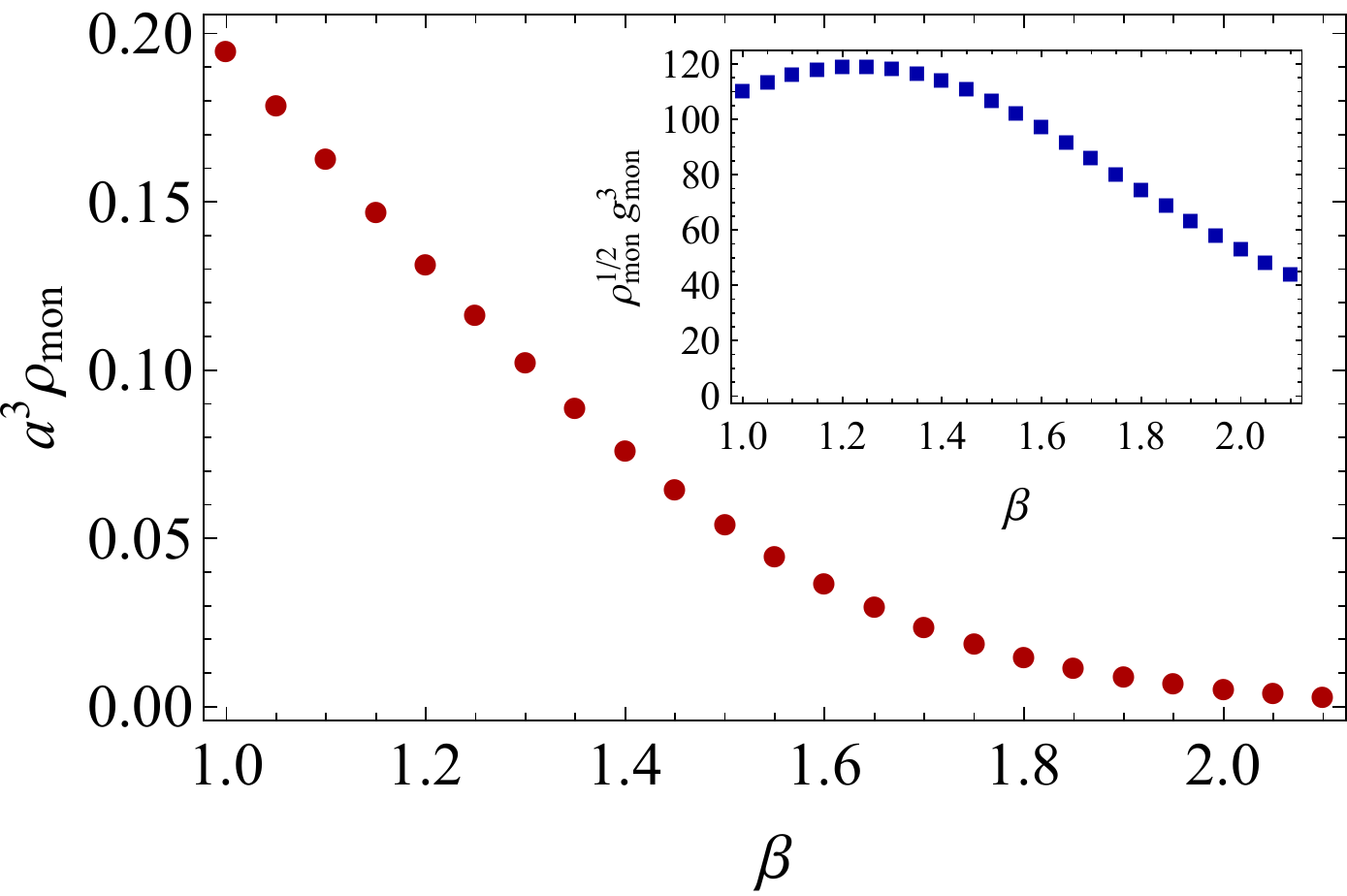}
\end{center}
\vskip -5mm 
\caption{The density of lattice monopoles (in lattice units) vs the lattice coupling $\beta$. The inset demonstrates the inapplicability of the 
dilute gas approximation~\eq{eq:applicability}.} 
\label{fig:monopole:density}
\end{figure}

In Fig.~\ref{fig:monopole:density} we show the monopole density $\rho_\mon^\lat = a^3 \rho_\mon$ (in lattice units) in the strong coupling region given by small values of $\beta$. Below we will be working in the region $\beta = 1 \sim 2$ where the monopole density is rather high and the monopole effects are expected to be rather strong. For comparison, at $\beta=3$ (the smallest coupling of our previous study~\cite{Chernodub:2016owp}) the density of monopoles in lattice units is $\rho = 1.8(1)\times 10^{-5}$ which means that on average we have less than one monopole at the whole $24^3$ lattice. The monopole effects were therefore very small at $\beta \geqslant 3$.

The inset in Fig.~\ref{fig:monopole:density} shows the quantity $\varrho^{1/2}_\mon g^3_m$ which is expected to be small in the dilute gas approximation~\eq{eq:applicability}. However, in the strong-coupling region this quantity is rather large indicating that the dilute gas approach should theoretically break down. Nevertheless we will see below that at relatively large $\beta$ (but still within the strong coupling region) certain features of the dilute monopole gas are manifestly visible.

\subsection{Effect of wires on monopoles}

\subsubsection{Monopole densities}

The Casimir effect and monopoles influence each other in both directions: the finite geometry imposed by the wires in the Casimir problem affects the dynamics of the monopoles while the presence of the dynamical monopoles modify the Casimir forces between the wires. In this section we discuss the former question and then in the next section we will address the latter. 

Firstly, let us make a comment on our terminology. The wires are one-dimensional objects in two spatial dimensions. Their world-surfaces are flat planes that are parallel to each other. The monopoles, in turn, are instanton-like objects, the positions of which are marked by the points in three-dimensional spacetime. Therefore it is more suitable to discuss the monopoles positioned in between the plates (the mentioned flat planes) rather than monopoles in between the wires. Below, we will alternatively speak about wires and plates (the latter ones are the world-surfaces of the wires themselves).

Secondly, let us discuss what could be an expected effect of the plates on the monopoles in between them? The plates affect the electromagnetic flux emanating from the volume confined in between the plates. In the perfect-metal limit, $\varepsilon \to \infty$, the plates should make the electromagnetic flux going into or out of the volume between the plate to vanish. According to the Gauss law the total magnetic charge density of monopoles should therefore be zero. Moreover, the closely-spaced plates should squeeze the spherical $3D$ configuration of magnetic flux around monopoles which should form a $2D$ configuration which is definitely more energetically costly.  Therefore we expect that the overall effect of the plates is likely to suppress the density of monopoles and antimonopoles in between the plates. One can also presume that the stronger permittivity $\varepsilon$ the stronger is the suppression of the monopole density. Finally, the larger distance between the plates $R$ the smaller the effect of the monopole suppression is expected be.  

\begin{figure*}[!thb]
\begin{center}
\vskip 3mm
\begin{tabular}{cc}
\includegraphics[scale=0.5,clip=true]{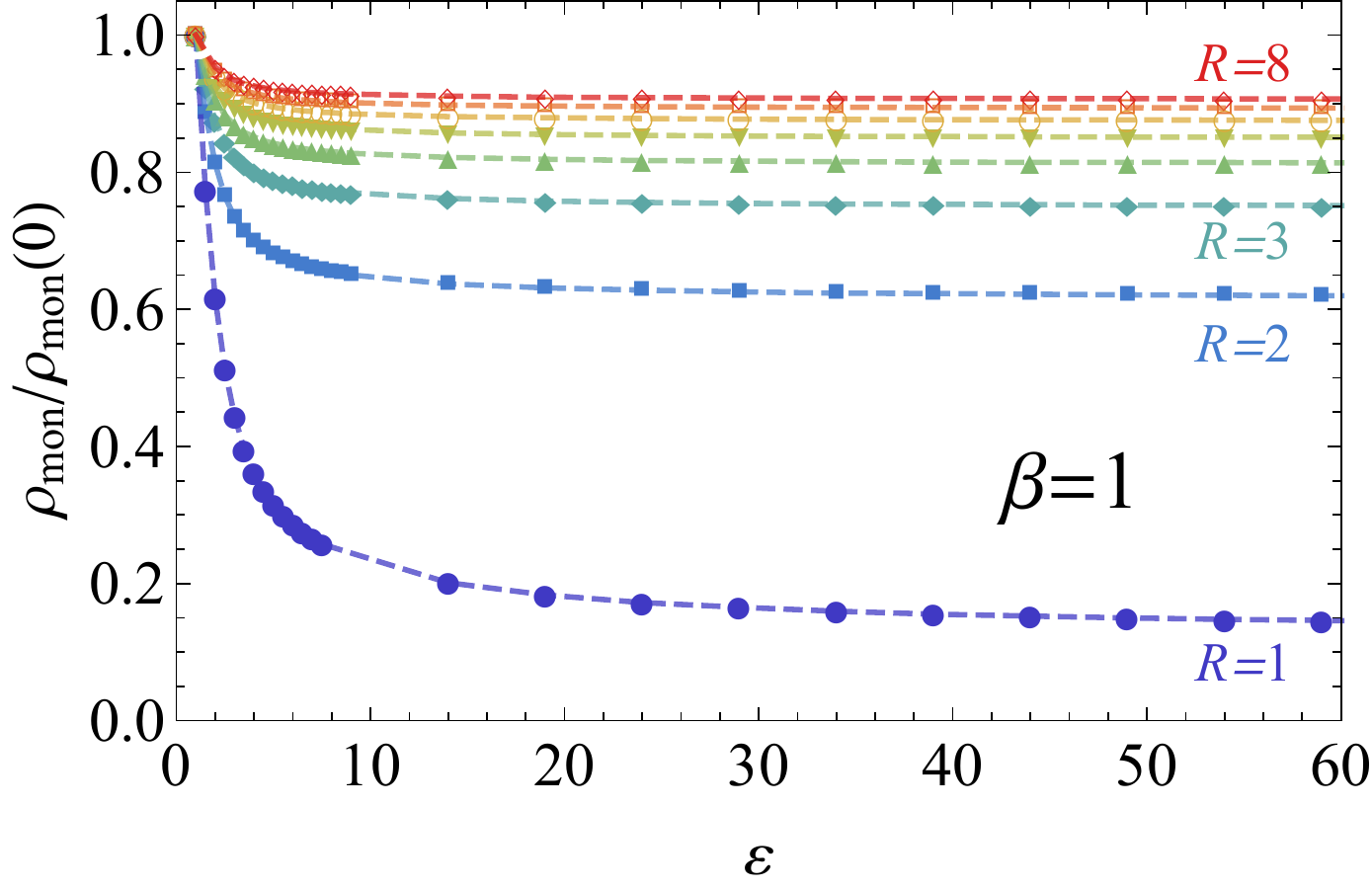} & \hskip 5mm
\includegraphics[scale=0.5,clip=true]{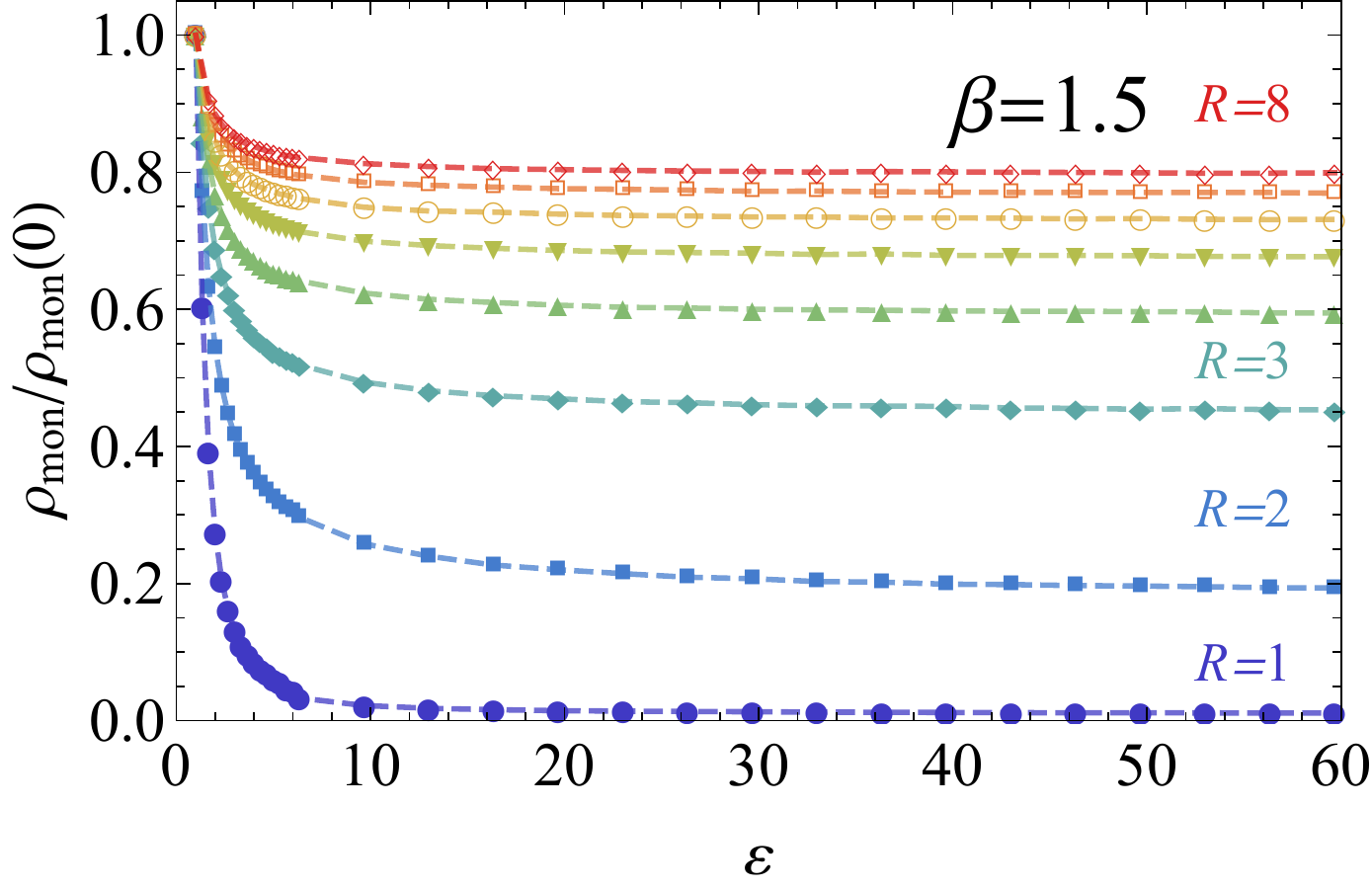}
\end{tabular}
\end{center}
\vskip -5mm 
\caption{Density of monopoles  $\rho_\mon$ in between the plates vs. permittivity $\varepsilon$ for various values of the lattice gauge coupling $\beta=1.0,1.1,1.2,1.3$ and distances between the plates $R = 1, 2, \dots, 8$ (in lattice units). The lines are the fits by the function~\eq{eq:fit:function}.} 
\label{fig:monopole:casimir}
\end{figure*}

All mentioned features are well seen in Fig.~\ref{fig:monopole:casimir} which shows the monopole density $\rho_\mon$ between the plates vs. permittivity $\varepsilon$ for various values of the lattice coupling $\beta$ and plate separation $R$. The closer the plates (wires) the smaller is the monopole density in between the wires. Increase of the permittivity results in diminishing of the monopole density. Qualitatively, all these properties are universal as they are largely independent of the lattice gauge coupling~$\beta$. The effect of the monopole suppression is especially strong at the smallest separation between the plates (wires), $R = a$.

\begin{figure}[!thb]
\begin{center}
\vskip 3mm
\includegraphics[scale=0.45,clip=true]{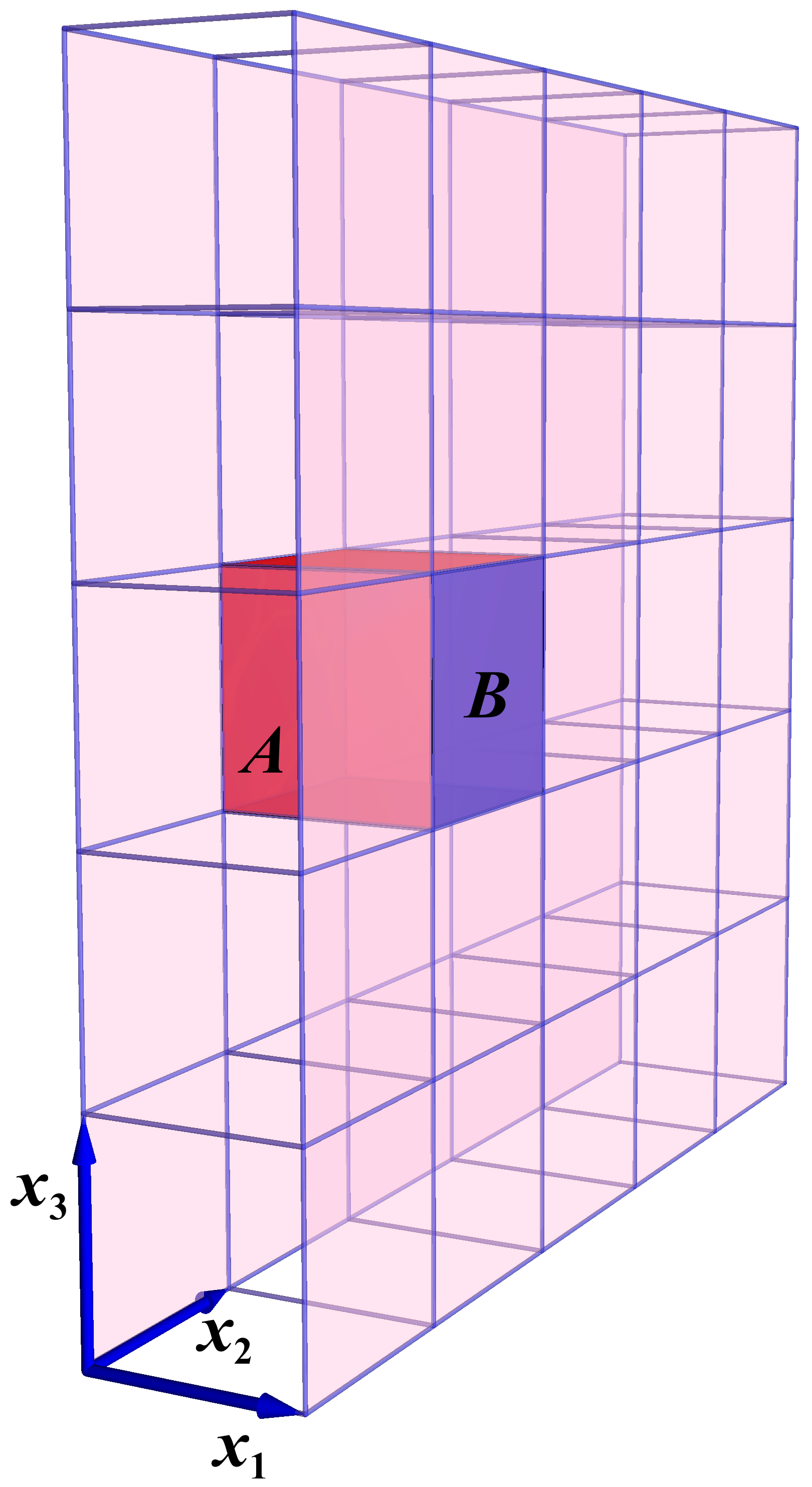}
\end{center}
\vskip -5mm 
\caption{Illustration of the monopole squeezed in between the plates (worldsheets of the wires) at the minimal distance $R=a$ between the wires.} 
\label{fig:monopole:squeezed}
\end{figure}

At the closest separation $R=1a$, the world-surfaces of the wires touch the two sides of a monopole in between the wires, Fig.~\ref{fig:monopole:squeezed}. At these sides, marked by ``$B$'' in the Figure, the magnetic flux is vanishing due to the lattice version~\eq{eq:F01:latt:3D} of the Casimir boundary condition~\eq{eq:F01:3d}. As the result, the magnetic flux of the monopole may only penetrate the into the subspace in between the plates via other four sides of the lattice monopole that are marked by ``$A$'' in Fig.~\ref{fig:monopole:squeezed}. The dynamics of the monopoles become, essentially, two-dimensional.

\begin{figure}[!thb]
\begin{center}
\vskip 3mm
\includegraphics[scale=0.55,clip=true]{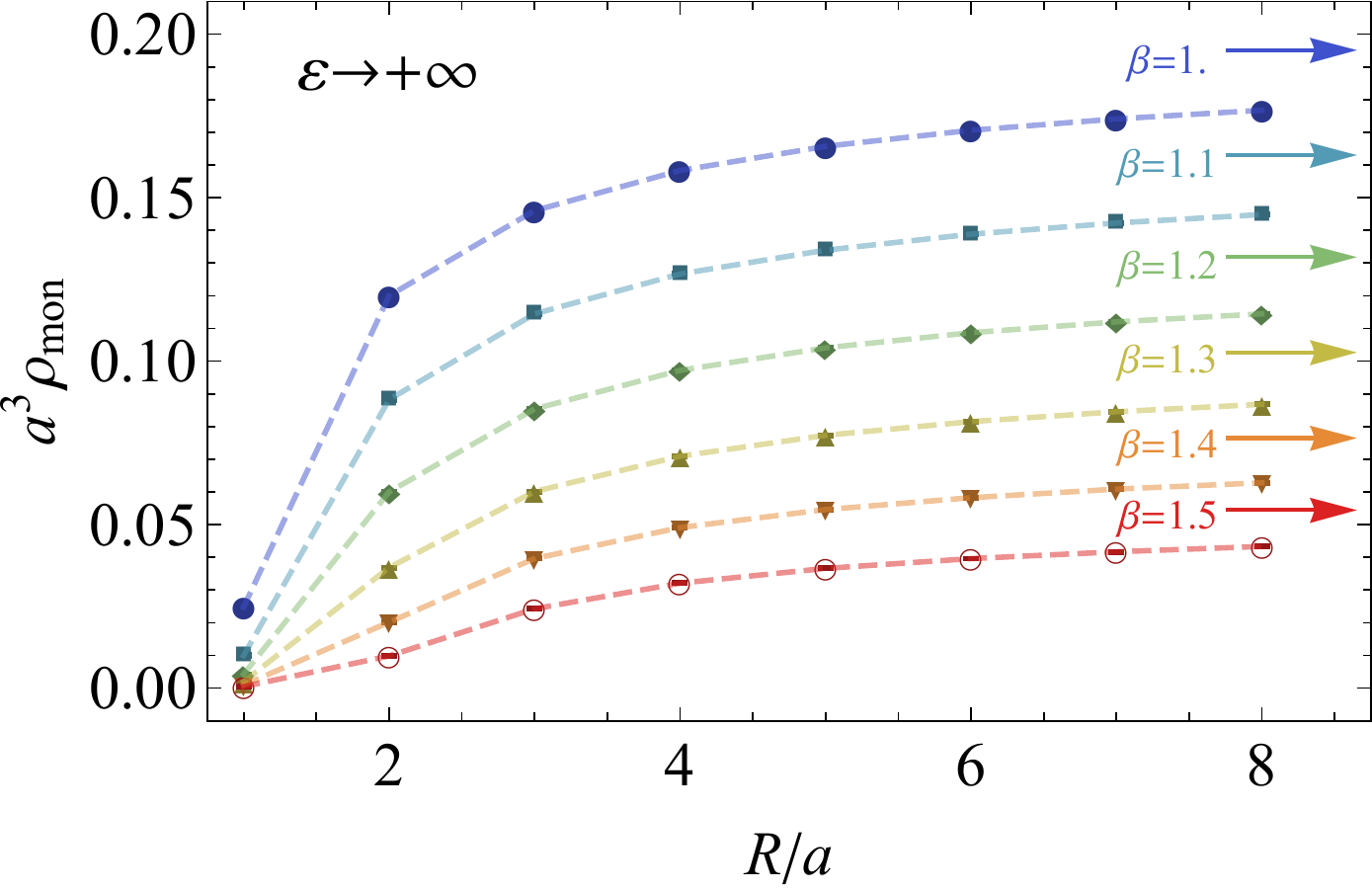}
\end{center}
\vskip -5mm 
\caption{Monopole density $\rho_{\mathrm{mon}}$ in between the wires as the function of the inter-wire distance $R$ (both are shown in the lattice units) extrapolated to the perfect metal limit for a set of fixed lattice gauge couplings $\beta$. The arrows show the asymptotic values $R \to \infty$ for each $\beta$. The latter are given by the monopole density in the absence of the wires (Fig.~\ref{fig:monopole:density}).} 
\label{fig:mon:dens:inside}
\end{figure}

According to Fig.~\ref{fig:monopole:casimir} the monopole density is a rather smooth function of the permittivity $\varepsilon$ at fixed coupling $\beta$ and inter-wire distance $R$. As we will see below, the $\varepsilon$  dependence of all quantities, including the monopole density, can be described with a very good accuracy by the following heuristic function:
\beqn
\cO^{\mathrm{fit}}(\varepsilon) = \cO_\infty + \frac{a_1}{\varepsilon + b_1} + \frac{a_2}{\varepsilon^2 + b_2}\,,
\label{eq:fit:function}
\eeqn
where the fitting parameters $a_{1,2}$ and $b_{1,2}$ control the slope of the $\varepsilon$ dependence while $ \cO_\infty$ corresponds to the value of the quantity $\cO$ in the limit $\varepsilon \to \infty$. Using Eq.~\eq{eq:fit:function} we extrapolate the monopole density (in this case $\cO \equiv \rho_{\mathrm{mon}}$) in between the plates to the perfect metal limit $\varepsilon \to \infty$. The result is shown in Fig.~\ref{fig:mon:dens:inside}.

Notice that in Ref.~\cite{Chernodub:2016owp} in a related investigation of the Casimir effect at weak coupling it was found that various expectation values can be excellently described the function~\eq{eq:fit:function} with three fitting parameters $\cO_\infty$, $a_1$ and $b_1$ while the second term in Eq.~\eq{eq:fit:function} may be put to zero with $a_2 = b_2 = 0$. At the strong coupling studied in the present paper we need both terms in Eq.~\eq{eq:fit:function} to successfully describe the data.

We see from Fig.~\ref{fig:mon:dens:inside} that in the perfect-metal limit $\varepsilon \to \infty$ the monopole density at the closest separation between the wires is zero. As the inter-wire distance $R$ increases the monopole density increases as well. The steepest slope of increase is achieved at stronger coupling $g$ [at smaller lattice coupling $\beta$, Eq.~\eq{eq:beta:3D}]. The latter fact is natural since in the absence of the wires the monopole medium is denser at stronger coupling $g$.

\subsubsection{Dimensional reduction and monopole transition}

What happens to the monopoles in between the plates? Surely, the approaching plates make the monopole density smaller due to squeezing of the magnetic flux and making monopoles heavier. This leads, as we discussed, to the overall suppression of the monopole density in between the plates. However, the very same effect of the restriction of the total magnetic flux of the monopoles from the three-dimensional space to the two-dimensional subspace, Fig.~\ref{fig:monopole:squeezed}, also affects the interactions between the monopoles and antimonopoles. In particular, the three dimensional attracting $1/r$ Coulomb potential~\eq{eq:D} transforms into a two dimensional logarithmic potential:
\beqn
D_{3D}({\bs x}) = \frac{1}{4 \pi | \bx |} \to D_{2D}({\bs x}) =  \frac{2}{R} \log \frac{| \bx |}{R} \,.
\label{eq:dim:red}
\eeqn

\begin{figure}[!thb]
\begin{center}
\vskip 3mm
\begin{tabular}{cc}
\includegraphics[scale=0.3,clip=true]{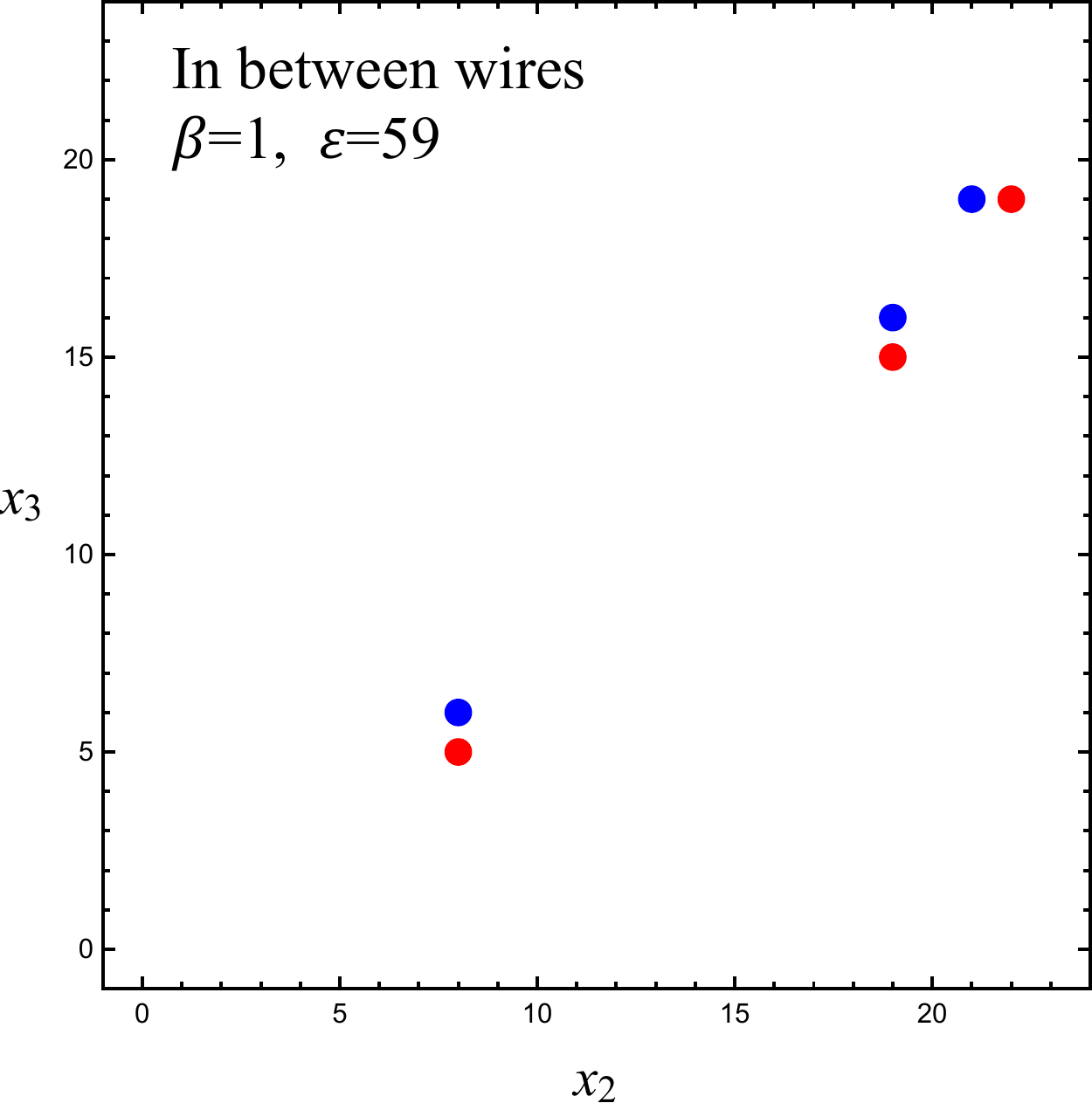} & 
\includegraphics[scale=0.3,clip=true]{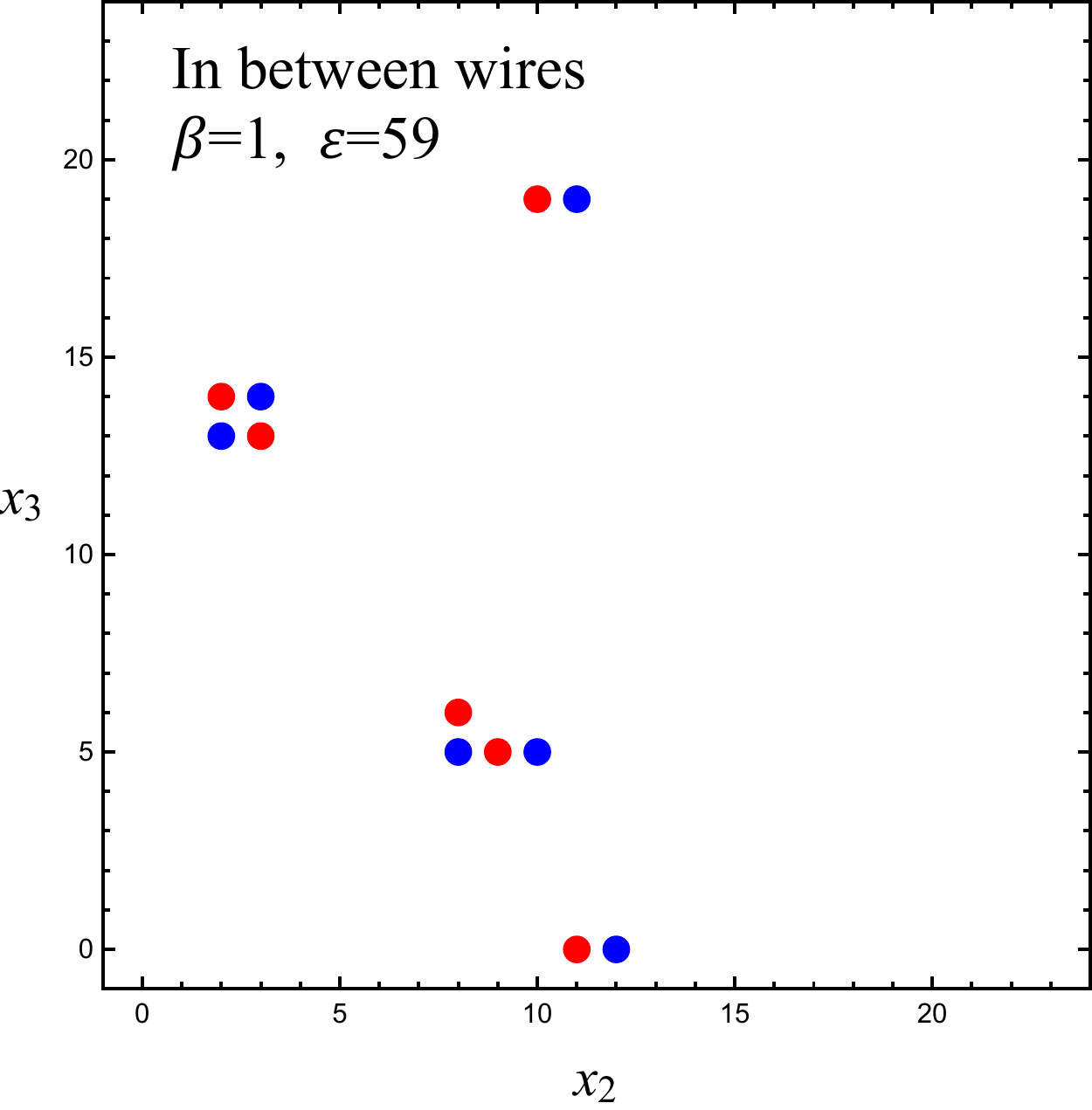} \\[3mm]
\includegraphics[scale=0.3,clip=true]{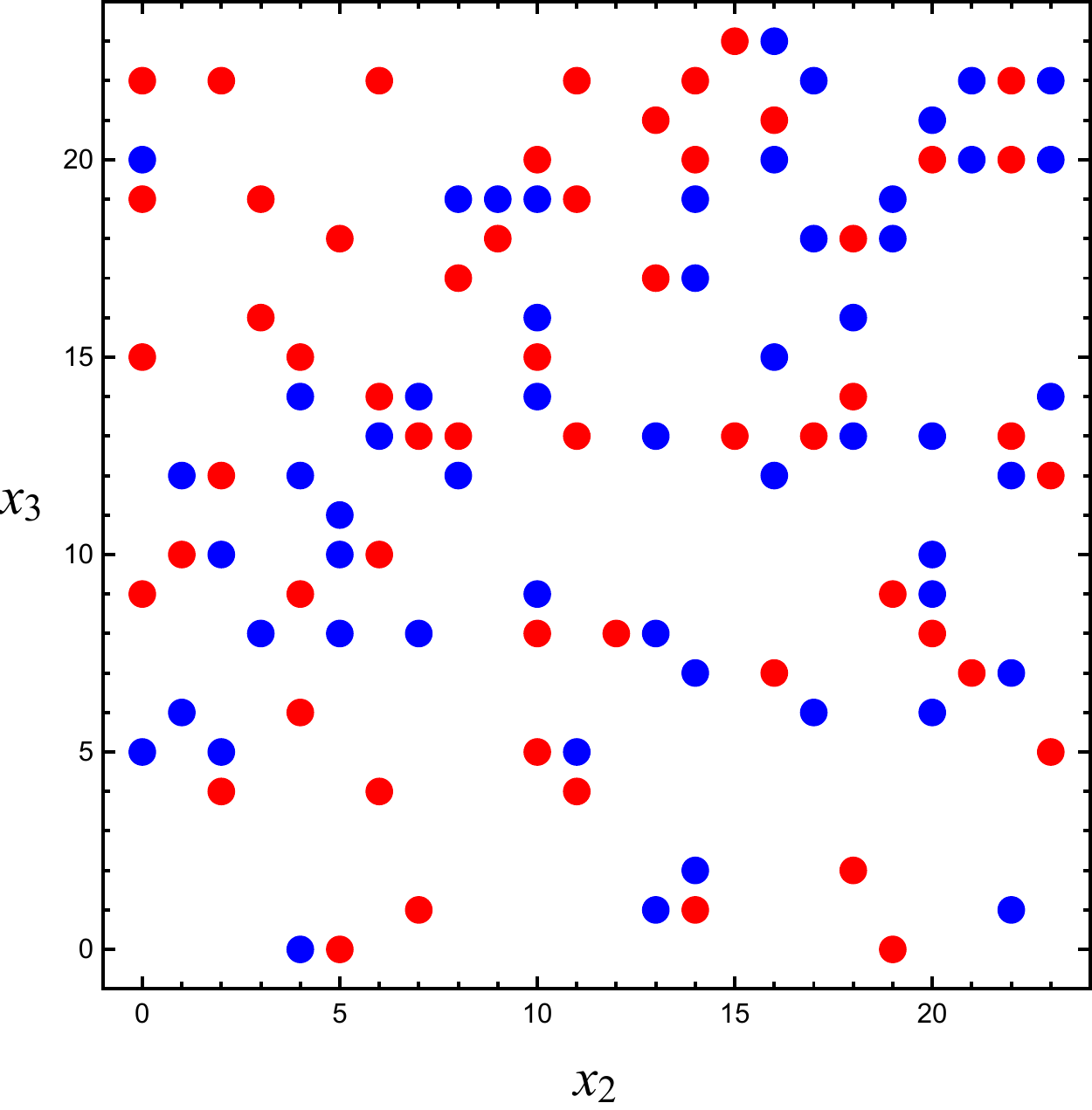} & 
\includegraphics[scale=0.3,clip=true]{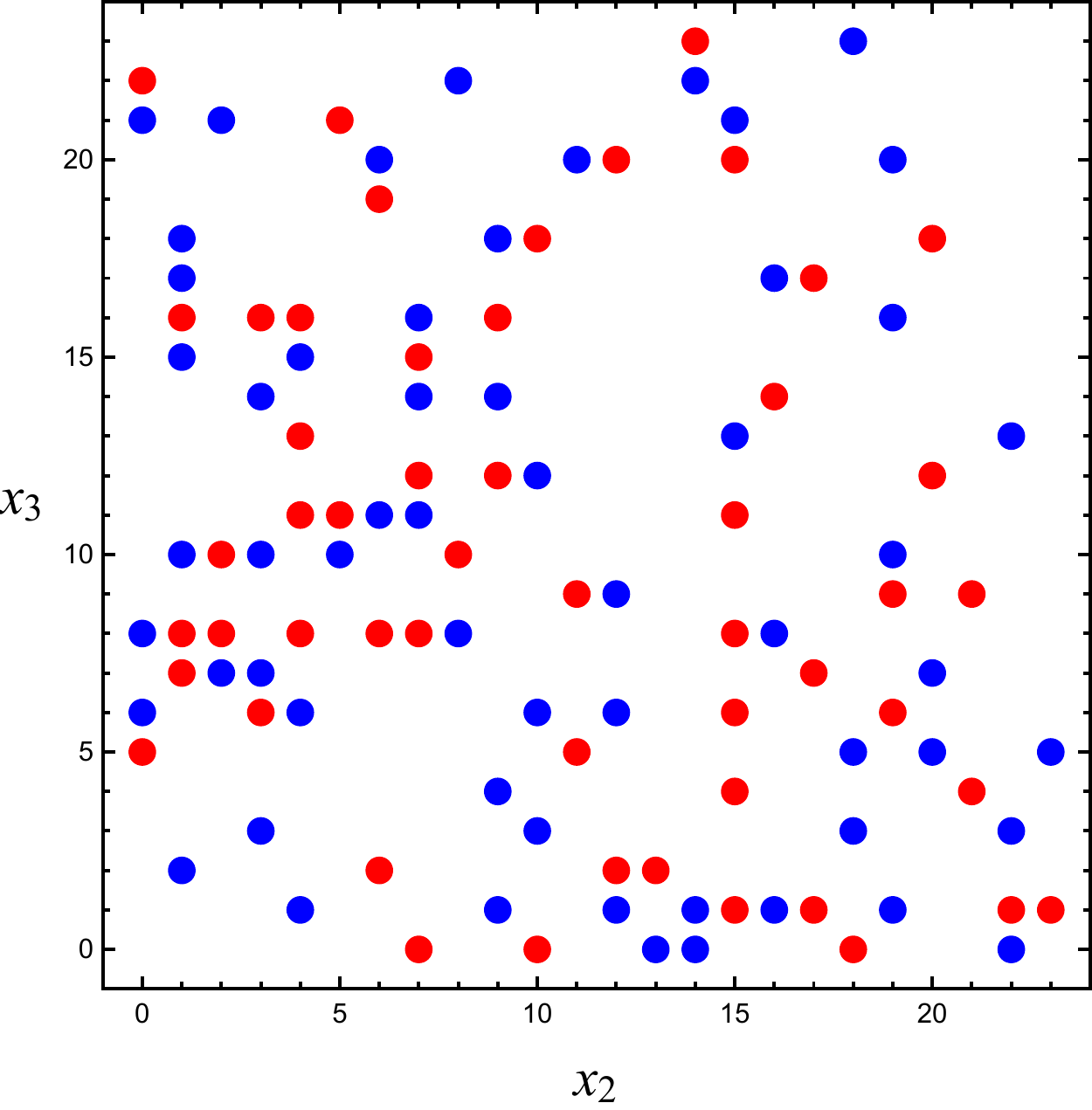} 
\end{tabular}
\end{center}
\vskip -2mm 
\caption{Examples of two configurations of monopoles (the red dots) and anti-monopoles (the blue dots) in the space between the wires (the upper plots) and outside the wires (the lower plots) at strong coupling regime ($\beta = 1$) and at large permittivity ($\varepsilon = 59$).}
\label{fig:monopoles}
\end{figure}

The dimensional reduction should have a strong effect on the monopole dynamics since the rapidly decreasing potential becomes the slowly rising logarithmic function~\eq{eq:dim:red}. Moreover, the logarithmic function represents a confining potential and therefore we expect that at sufficiently large values of the permittivity $\varepsilon$ the individual monopoles should be confined into the magnetic dipole pairs that consist of closely spaced monopoles and anti-monopoles. This transition is indeed seen in our numerical simulations: in Figure~\ref{fig:monopoles} we visualize typical monopole configurations in between and outside the wires for a large value of permittivity $\varepsilon$. At small values of  $\varepsilon$ the monopoles in between the wires resemble a gas of monopoles rather than the gas of magnetically neutral dipoles. Thus the increasing permittivity leads to a transition of the monopole gas into the dipole gas in the space between the wires.

The transition from monopole gas to the magnetic dipole gas should lead to the absence of the mass-gap generation and to a confinement-deconfinement transition in the region between the plates. As we discuss later, the absence of the mass gap should give rise to a certain enhancement of the zero-point potential between the wires. Moreover, one could expect on general grounds that the physical picture should be similar to the finite-temperature deconfining transition in three-dimensional compact electrodynamics~\cite{Chernodub:2001ws,Chernodub:2000uv} . However, in our numerical simulations we found no conclusive evidence of the suggested phase transition from the confinement to deconfinement phases. In order to study the deconfinement we have calculated the vacuum expectation value of the Polyakov loop 
\beqn
L(x_1,x_2) = \exp\left\{i \sum_{x_3=0}^{L_t -1} \theta_{3}(x_1,x_2,x_3)\right\}\,,
\label{eq:Polyakov:Loop}
\eeqn
inside and outside the plates. This quantity is the order parameter of confinement: it is zero in the confinement phase and nonzero otherwise. 

We found that at zero temperature the Polyakov loop is consistent with a zero value inside and outside plates. The absolute values of the bulk sum of the Polyakov loop~\eq{eq:Polyakov:Loop} over all the points inside and, separately, outside of the plates do not show the existence of the phase transition at all studied values of parameters (coupling constants $\beta$, permittivity $\varepsilon$ and interwire separations $R$). Most plausibly this negative result comes due to the fact that the Polyakov loop is not a convenient variable to study confinement of charges at zero temperature. We are planning to readdress this question at finite-temperature theory in our next study. 

\subsection{Monopoles, energy and pressure}

\subsubsection{Stress energy tensor on the lattice}

The zero-point energy of the quantum field fluctuations is affected by the vicinity of the wires. Since the zero-point (Casimir) energy depends on the interwire distance $R$, the modification of the vacuum fluctuations leads to appearance of a force applied to the wires. This phenomenon is known as the Casimir effect. 

The Casimir effect can be calculated by a direct evaluation of the energy density of vacuum fluctuations which is given by is the ``temporal-temporal'' component $T^{00}$ of the  Lorentz-covariant energy-momentum (stress\--energy) tensor of the gauge field~\eq{eq:S:continuum},
\beqn
T^{\mu\nu} = - \frac{1}{g^2}F^{\mu\alpha} F^\nu_{\ \alpha} + \frac{1}{4 g^2} \eta^{\mu\nu} F_{\alpha\beta} F^{\alpha\beta}.
\label{eq:T:munu}
\eeqn

In Minkowski spacetime the diagonal components of the energy-momentum tensor~\eq{eq:T:munu} of the gauge field are as follows:
\begin{subequations}
\beqn
T^{00} & = & \frac{1}{2 g^2} \left(E_x^2 + E_y^2  + B_z^2\right),
\label{eq:T00:M} \\
T^{11} & = & \frac{1}{2 g^2} \left(- E_x^2 + E_y^2  + B_z^2\right),
\label{eq:T11:M} \\
T^{22} & = & \frac{1}{2 g^2} \left(E_x^2 - E_y^2  + B_z^2\right).
\label{eq:T22:M} 
\eeqn
\label{eq:T:munu:M}
\end{subequations}
\!\!In the spacetime with the metric $(+,-,-)$ the components of the field-strength tensor~\eq{eq:F:munu} are $F_{01} = E_x$,  $F_{02} = E_y$ and $F_{12} = - B_z$. The former two ones are the components of the electric fields acting along $x$ and $y$ axis, while the remaining component $B_z$ has a formal sense of the magnetic field although there is no $z$ axis in the (2+1) dimensional Minkowski spacetime.

In Euclidean spacetime the components~\eq{eq:T:munu:M} of the energy-momentum tensor are as follows:
\begin{subequations}
\beqn
T^{00}_E & = & \frac{1}{2 g^2} \left( - E_x^2 - E_y^2 + B_z^2\right),
\label{eq:T00:E} \\
T^{11}_E & = & \frac{1}{2 g^2} \left( - E_x^2 + E_y^2 - B_z^2\right),
\label{eq:T11:E} \\
T^{22}_E & = & \frac{1}{2 g^2} \left( E_x^2 - E_y^2 - B_z^2\right),
\label{eq:T22:E} 
\eeqn
\label{eq:T:munu:E}
\end{subequations}
\!\!\!\!where we took into account that as we pass from Minkowski to Euclidean spacetime the terms with electric field in Eq.~\eq{eq:T:munu:M} change their signs, $E_x^2 \to - E_x^2$  and $E_y^2 \to - E_y^2$, while the magnetic field remains intact, $B_z^2 \to B_z^2$. Moreover, $T^{00}\to T^{00}_E$ while $T^{11}\to -T^{11}_E$  and $T^{22}\to - T^{22}_E$ .

At zero temperature the system is invariant under discrete rotations by the angle $\pm \pi/2$ around the $x \equiv x_1$ axis. Since after these rotations the fields $E_x$ and $B_z$ are interchanged, one has $E_x \leftrightarrow \pm B_z$, and, consequently,
\beqn
\avr{E_x^2} = \avr{B_z^2}\,.
\label{eq:invariance}
\eeqn
Therefore we conclude from Eqs.~\eq{eq:T:munu:E} and \eq{eq:invariance} that
\beqn
\avr{T^{00}_E} & = & \avr{T^{22}_E} = - \frac{1}{2 g^2} \avr{E_y^2}\,,  
\label{eq:T00:T22:E}\\
\avr{T^{11}_E} & = & \frac{1}{2 g^2} \bigl(\avr{E_y^2} - 2 \avr{E_x^2}   \bigr)\,.
\label{eq:T11:E:2}
\eeqn

In Minkowski spacetime the temporal component $T^{00}$ of the strength-energy tensor has the sense of the energy density while the spatial components $T^{11}$ and $T^{22}$ are the stresses that are usually associated with the (local) pressure. Notice that in the presence of the closely spaced wires the directions $x \equiv x_1$ and $y \equiv x_2$ are not equivalent, and therefore the local stresses in these directions may be different. Making the Wick rotation to the Euclidean spacetime, we get that the local (stress) pressure along the wires is equal to the energy density~\eq{eq:T00:T22:E} while it is not equal to the stress in the transversal direction with respect to the wires~\eq{eq:T11:E:2}.

In the ultraviolet limit the expectation value of the energy density~\eq{eq:T00:T22:E} is a divergent quantity both in the presence and in the absence of the wires. The difference between these expectation values is the physical, ultraviolet-finite quantity which has a sense of an excess in the energy density of quantum fluctuations due to the presence wires. Therefore the physical (measurable) effect of the wires can be quantified in terms of the normalized energy density:
\beqn
{\cal E}_R(x) & 
= & \avr{T^{00}_E(x)}_{R} - \avr{T^{00}_E(x)}_{0},
\label{eq:E:norm}
\eeqn
where the subscripts ``$R$'' and ``0'' indicate that these expectation values are taken, respectively, in the presence of the wires separated by the distance $R$ and in the absence of the wires.

Since the wires are parallel to each other the energy density~\eq{eq:E:norm} depends only on the coordinate $x_1$, which is transversal to the wires themselves. Therefore it is natural to introduce the total (Casimir) energy density with respect to the unit length of the wires:
\beqn
V_{\Cas}(R) = \int\limits_{-\infty}^{+\infty} d x_1\, {\cal E}_R(x_1) \equiv - \frac{1}{2 g^2} \aavr{E_y^2}\,,
\label{eq:V:Cas}
\eeqn
where we used Eq.~\eq{eq:T00:T22:E} and, following Ref.~\cite{Chernodub:2016owp}, introduced the average
\beqn
\avr{\!\avr{\cO(x)}\!} = \int d x_1 \left[ \avr{\cO(x)}_R - \avr{\cO}_0 \right]\,,
\label{eq:avr:O}
\eeqn
which corresponds to the excess of the expectation value of the quantity $\cO$ evaluated per unit length of the wires. In the lattice regularization the average~\eq{eq:avr:O} has the following form:
\beqn
\avr{\!\avr{\cO(x)}\!}_{\lat} =\sum_{x_1 = 0}^{L_s - 1} \left[ \avr{\cO(x_1)}_R - \avr{\cO}_0 \right]\,.
\label{eq:avr:O:lat}
\eeqn
Thus, the lattice Casimir energy density per unit length of the wires~\eq{eq:V:Cas} takes the following compact form:
\beqn
V_{\Cas}(R) = \beta \aavr{\cos \theta_{23}},
\label{eq:V:Cas:lat}
\eeqn
where the lattice gauge coupling $\beta$ is given in Eq.~\eq{eq:beta:3D}. Here we naturally redefined the Casimir energy in the lattice units, $V_{\Cas} \to a^2 V_{\Cas}$.

For our purposes it is also convenient to consider an unnormalized version of the Casimir potential 
\beqn
V^w_{\Cas}(R) = \beta \avr{\cos \theta_{23}}_R.
\label{eq:V:Cas:lat:unnorm}
\eeqn

Similarly to the energy density~\eq{eq:V:Cas:lat}, one can also define the (cummulative) transversal pressure of the quantum fluctuations,
\beqn
P_{x}(R) = \beta \bigl(2 \aavr{\cos \theta_{13}} - \aavr{\cos \theta_{23}} \bigr),
\label{eq:P:Cas:lat}
\eeqn
which is associated with the $T^{11}$ component of the energy-momentum tensor~\eq{eq:T11:E:2}. The quantity~\eq{eq:P:Cas:lat} is the energy-momentum stress in the $x_1$ direction integrated over the transverse spatial direction in a manner of Eq.~\eq{eq:V:Cas}.

It is important to notice that the pressure experienced by the wires due to the zero-point fluctuations is not given by the integrated value of the $T^{11}$ stress~\eq{eq:P:Cas:lat}. Naturally, the pressure experienced by each wire is equal to the difference of the values of the $T^{11}$ component of the energy-momentum tensor at inner and outer sides of the wire: 
\beqn
P_\pm = T^{11}(\pm R/2 + \epsilon) - T^{11}(\pm R/2 - \epsilon)\,,
\eeqn
where $\epsilon \to 0$ is a vanishing positive quantity. It turns out that~\cite{ref:Milton}
\begin{itemize}
\item[(i)] the pressure which is experienced by the left wire $P_-$ is opposite in sign with respect to the pressure $P_+$ which is felt by the right wire, $P_+ = - P_-$; 

\item[(ii)] the force of attraction, that is pressure times length of each wire, $F = P L$, is equal to the minus derivative of the Casimir energy, 
\beqn
F(R) = - E'_\Cas(R)\,.
\label{eq:F:R}
\eeqn
\end{itemize}

Summarizing, the energy density of the zero-point fluctuations~\eq{eq:T00:M} and their pressure (cumulative normal stress) in the direction normal to the wires~\eq{eq:T11:M},  integrated over the separation between the wires, are given by the lattice expressions in, respectively, Eq.~\eq{eq:V:Cas:lat} and Eq.~\eq{eq:P:Cas:lat}. Belo we study their properties numerically.

\subsubsection{Casimir energy and monopoles}

\begin{figure}[!thb]
\begin{center}
\vskip 3mm
\includegraphics[scale=0.55,clip=true]{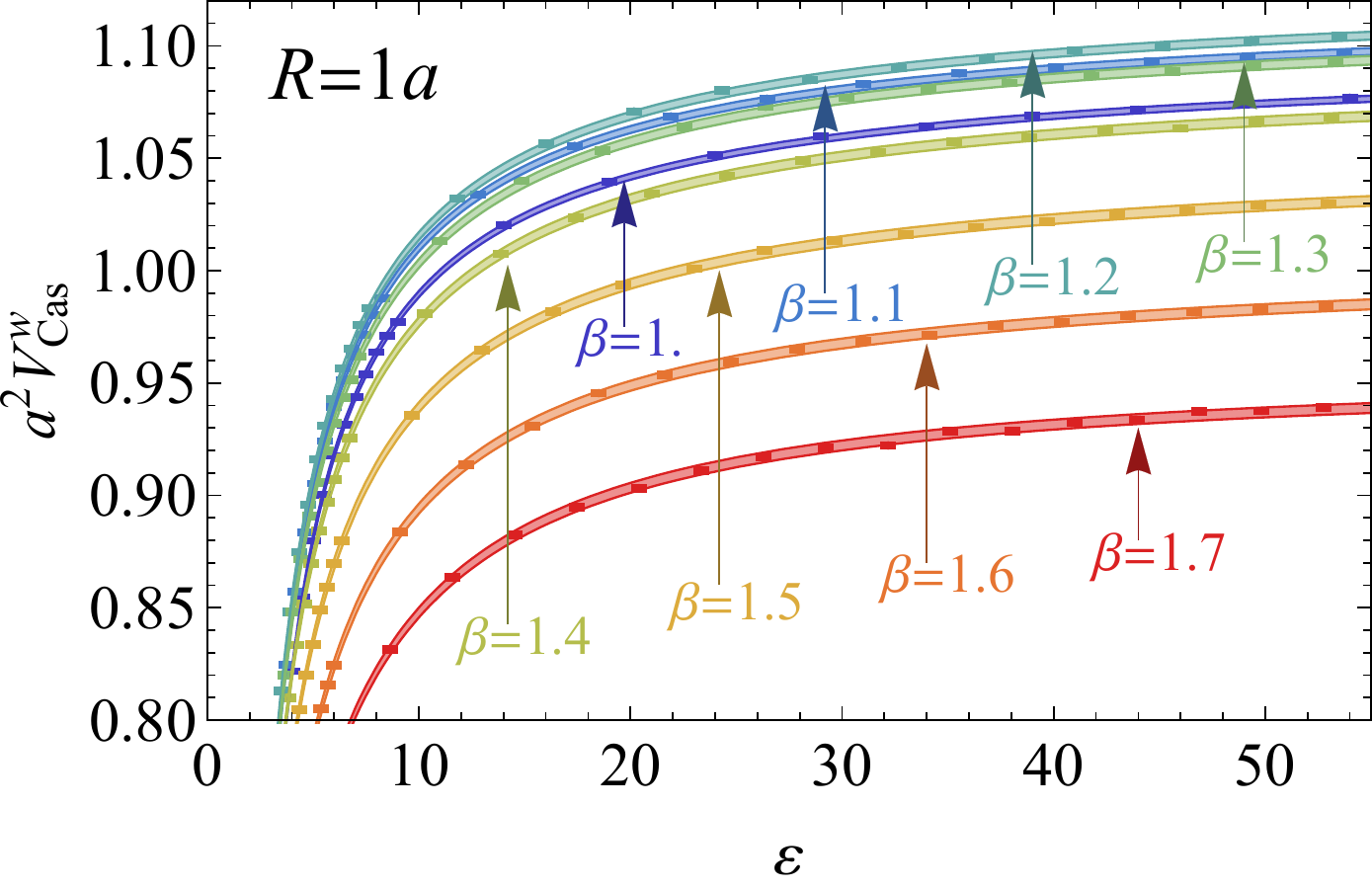}
\end{center}
\vskip -5mm 
\caption{The unnormalized lattice Casimir potential~\eq{eq:V:Cas:lat:unnorm} as function of permittivity $\varepsilon$ at a set of lattice couplings $\beta$. The extrapolation to the perfect-metal limit, shown by the continuous lines, is done by the fitting function~\eq{eq:fit:function}. The thickness of the lines corresponds to the errors of extrapolation.} 
\label{fig:Vcas:extrapolation}
\end{figure}

Our numerical data indicates that the Casimir potential is a smooth function of the wire permittivity $\varepsilon$. In Figure~\ref{fig:Vcas:extrapolation} we show the raw data unnormalized potential~\eq{eq:V:Cas:lat:unnorm} for the minimal separation between the wires, $R = 1a$ and for various values of the lattice coupling $\beta$. It is clearly seen that the increase in the permittivity $\varepsilon$ of the wires leads to the increase of the vacuum energy for all values of $\beta$. This property is quite natural since as at a higher value of  $\varepsilon$ the wires are more ``visible'' to the vacuum fluctuations compared to the lower-$\varepsilon$ wires.

Following the general strategy of Ref.~\cite{Chernodub:2016owp} we extrapolate the potential to the ideal metal limit $\varepsilon \to \infty$ by fitting the finite-$\varepsilon$ data by the function~\eq{eq:fit:function}. Due to the nonlinear nature of the function~\eq{eq:fit:function} we fit separately both terms~\eq{eq:avr:O:lat} in the normalized potential~\eq{eq:V:Cas:lat}. The fits of the first term in Eq.~\eq{eq:avr:O:lat} corresponding to the unrenormalized potential~\eq{eq:V:Cas:lat:unnorm} are also shown in Fig.~\ref{fig:Vcas:extrapolation}.

\begin{figure}[!thb]
\begin{center}
\vskip 3mm
\includegraphics[scale=0.55,clip=true]{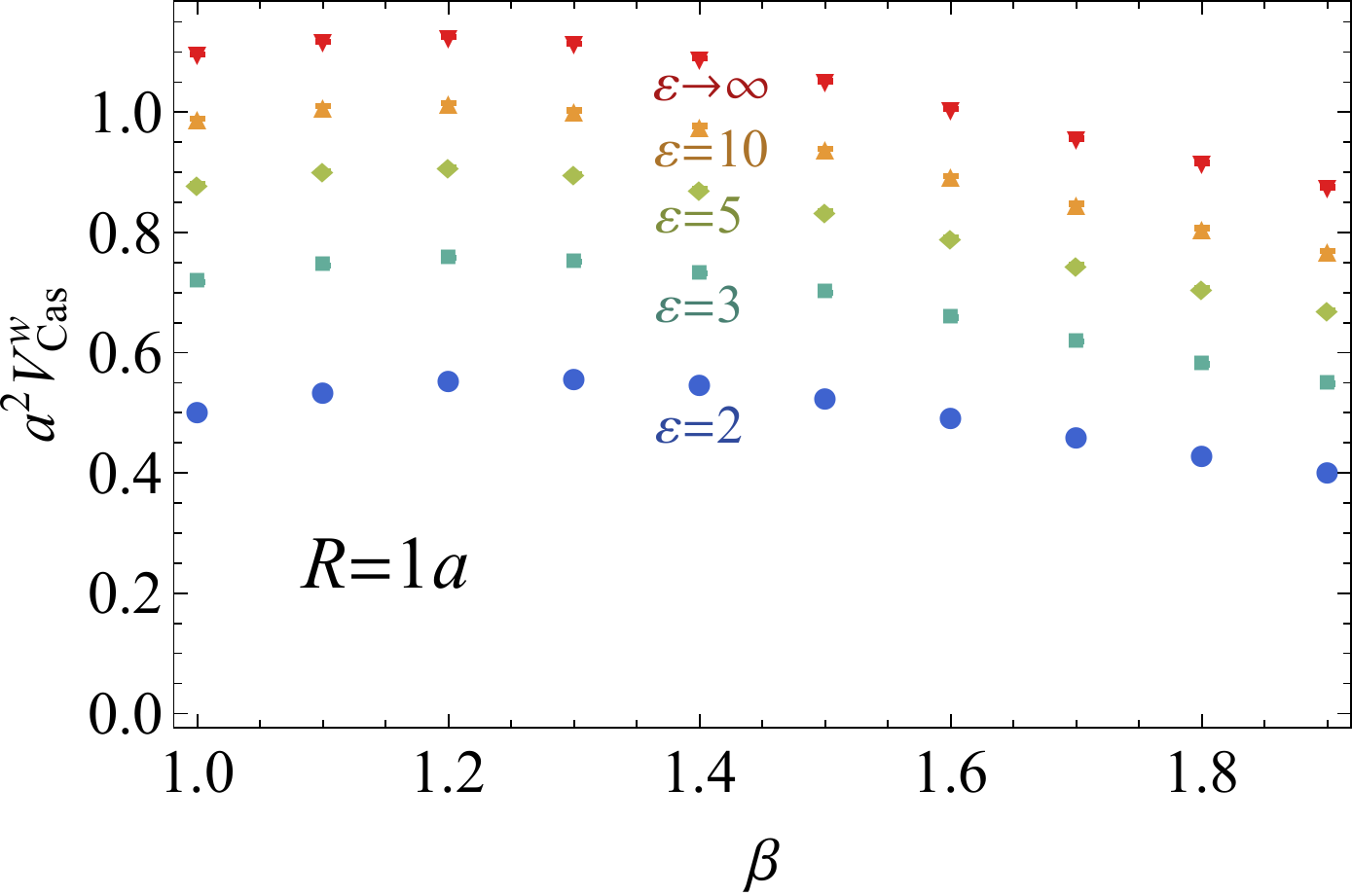}\\
(a) \\[5mm]
\includegraphics[scale=0.55,clip=true]{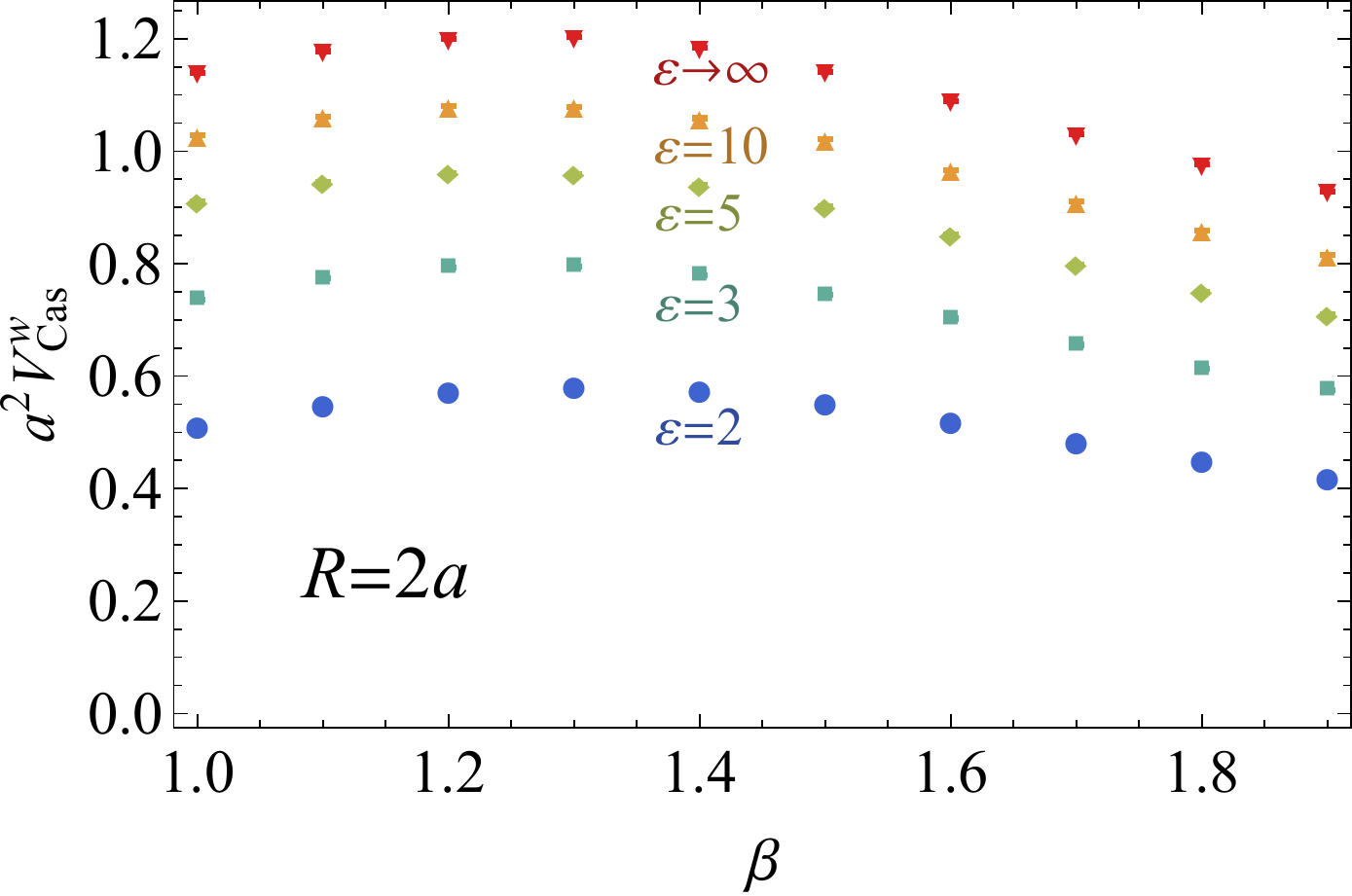}\\
(b)
\end{center}
\vskip -5mm 
\caption{The unnormalized Casimir potential~\eq{eq:V:Cas:lat:unnorm} vs. the lattice coupling $\beta$ at interwire separations $R=1a$ (the upper plot) and $R=2a$ (the lower plot) and various values of permittivity $\varepsilon$.} 
\label{fig:Vcas:R1:R2}
\end{figure}

In Figure~\ref{fig:Vcas:R1:R2} we show the (unnormalized) Casimir potential~\eq{eq:V:Cas:lat:unnorm} as a function of the lattice coupling constant $\beta$ at two interwire separations. We see that the larger permittivity $\varepsilon$ the larger is the (unrenormalized) Casimir potential and the stronger is the effect of the wires on vacuum fluctuations. The unrenormalized Casimir potential is a non-monotonic function of the coupling constant $\beta$ with $\varepsilon$-dependent positions of maxima. As the permittivity $\varepsilon$ increases, the maximum of the unrenormalized vacuum energy shifts towards smaller values of $\beta$. In the $\varepsilon \to \infty$ limit, which is obtained with the help of function~\eq{eq:fit:function}, the maximum of the unnormalized Casimir energy is achieved at $\beta \simeq 1.2$. At approximately the same value of the coupling constant the physical density of the monopoles achieves its maximum as it is shown in the inset of Fig.~\ref{fig:monopole:density}. 

\begin{figure}[!thb]
\begin{center}
\vskip 3mm
\includegraphics[scale=0.55,clip=true]{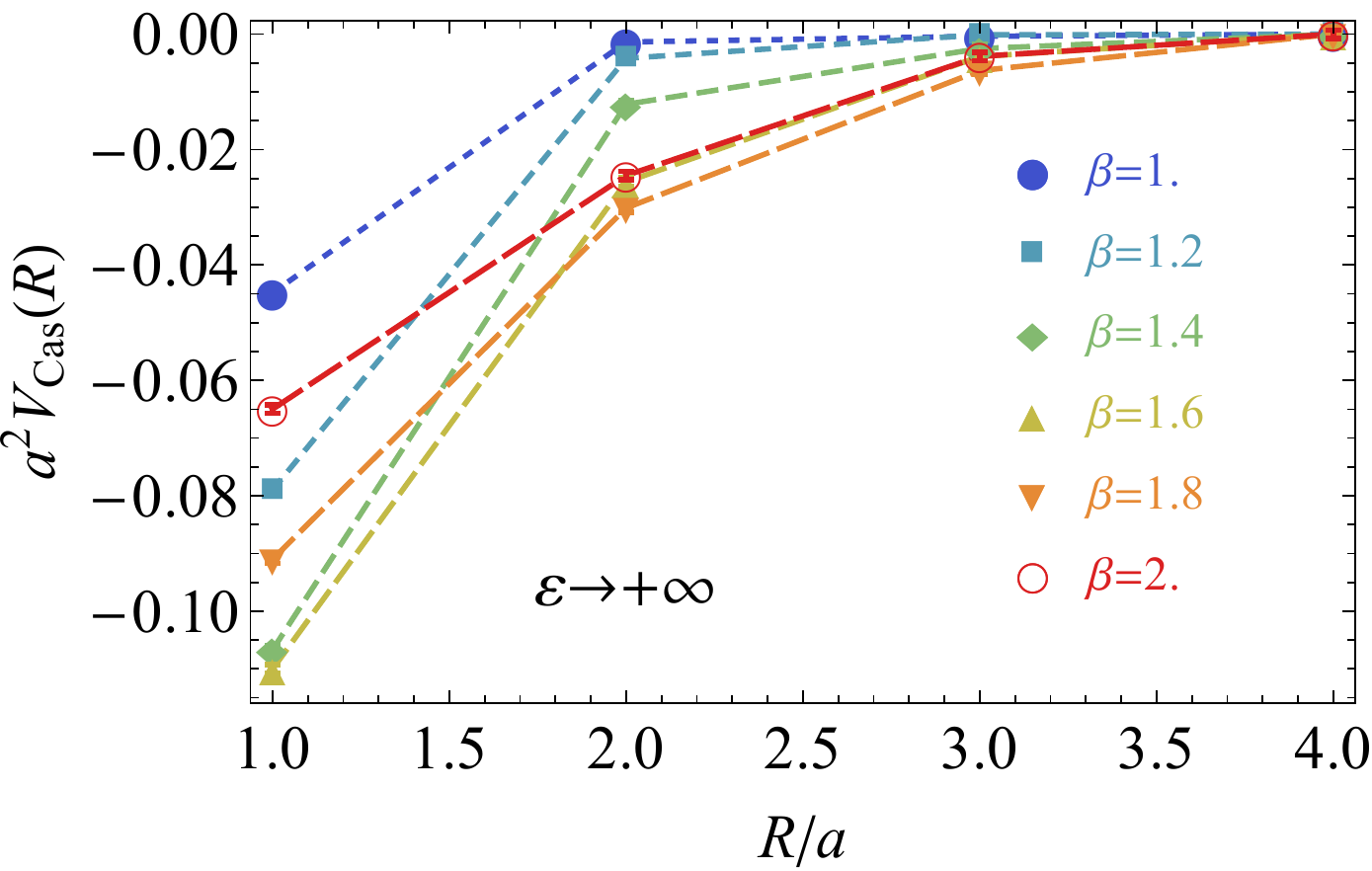}
\end{center}
\vskip -5mm 
\caption{The physical (normalized) Casimir potential~\eq{eq:V:Cas:lat} vs. the interwire distance $R$ at various values of the lattice coupling constant~$\beta$ (in lattice units).} 
\label{fig:Vcas:lattice}
\end{figure}

The physical (normalized) Casimir potential is a monotonic function of the distance $R$. In Figure~\ref{fig:Vcas:lattice} we show the potential in the limit of infinite permittivity (at finite permittivity $\varepsilon$ the potential resembles the limiting case $\varepsilon \to 0$ albeit smaller values of the amplitude). For all studied values of the couplings and distances the potential is a negative quantity. From Fig.~\ref{fig:Vcas:lattice} we can deduce a few interesting features of the Casimir energy:

\begin{enumerate}

\item At a weaker gauge coupling, at the lattice coupling $\beta \simeq 2$, the potential has a moderate strength and it is relatively long-ranged in similarity with our previous results~\cite{Chernodub:2016owp}.

\item In the stronger gauge coupling regime, as $\beta$ decreases, the potential becomes more short-ranged while its strength at short distances increases. 

\item As the coupling becomes even stronger and the lattice coupling approaches $\beta = 1$, the potential becomes even more short-ranged and the strength of the potential decreases. 

\end{enumerate}

Most these numerical findings may be understood from our analytical results obtained in Sect.~\ref{sec:compact:QED:monopoles} in the dilute gas approximation to the dynamics of the monopoles. Indeed, the monopoles lead to the mass gap generation which results in the finite mass of photon~\eq{eq:m:ph:rho}. As the photon becomes massive, the Casimir effect becomes naturally smaller and its effective radius of interaction decreases, as the Casimir interaction between the wires gets an additional exponential factor according to Eqs.~\eq{eq:V:Cas:R:massive}, \eq{eq:f:x} and \eq{eq:f:mon:expansion}. Since the interaction is very short-ranged we were not able to fit the numerically obtained results by the theoretical formula given in Eqs.~\eq{eq:V:Cas:R:massive} and \eq{eq:f:x}. 

In addition, our theoretical analysis indicates that the numerically observed enhancement of the Casimir potential at intermediate coupling cannot be explained by simple analytical calculations in the dilute gas approximation. However, one may strongly believe that the formation of the monopole-antimonopole pairs in between the plates -- discussed earlier and illustrated in Fig.~\ref{fig:monopoles} -- makes the Casimir potential long-ranged due to the absence of the mass gap. On the other the increased density of the monopoles outside the plates contributes to the pressure and increase the Casimir potential. 

\begin{figure}[!thb]
\begin{center}
\vskip 3mm
\includegraphics[scale=0.55,clip=true]{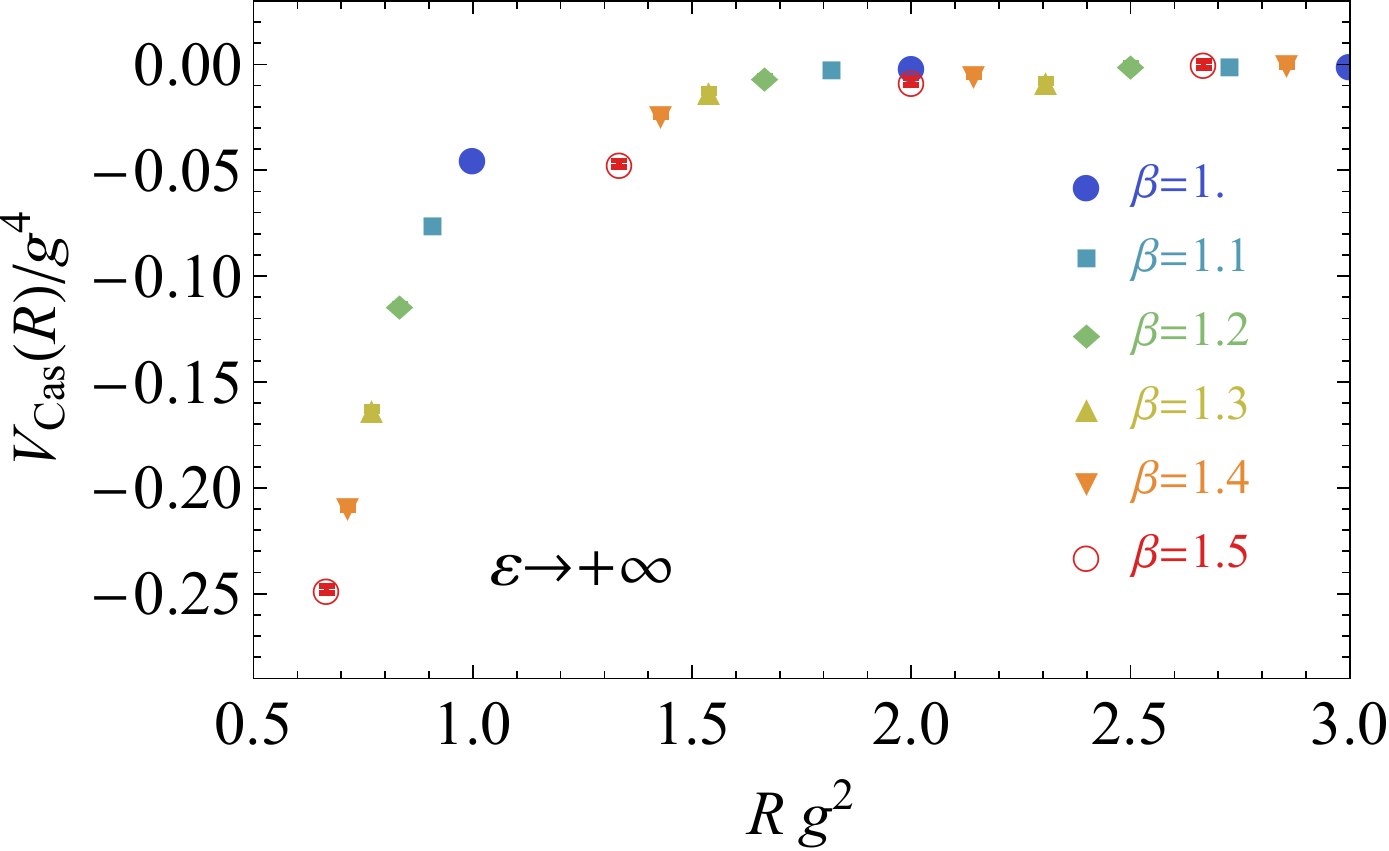}
\end{center}
\vskip -5mm 
\caption{Absence of scaling of the Casimir potential~\eq{eq:V:Cas:lat} in the strong coupling regime. All quantities are shown in physical units.} 
\label{fig:Vcas:scaling}
\end{figure}

We would like to notice that in the strong coupling regime there is no physical scaling of the Casimir potential contrary to the weak coupling regime considered in Ref.~\cite{Chernodub:2016owp}. This is the expected property because the lattice formulation of the compact electrodynamics represents inherently lattice gauge theory defined at the specific cutoff $a$. For example the relation~\eq{eq:beta:3D}  between the physical gauge coupling $g$ and the lattice spacing $a$ is valid only in the weak coupling regime. We demonstrate the absence of the formal scaling in Fig.~\ref{fig:Vcas:scaling}. The results are shown in the limit $\varepsilon \to \infty$.

\begin{figure}[!thb]
\begin{center}
\vskip 3mm
\includegraphics[scale=0.55,clip=true]{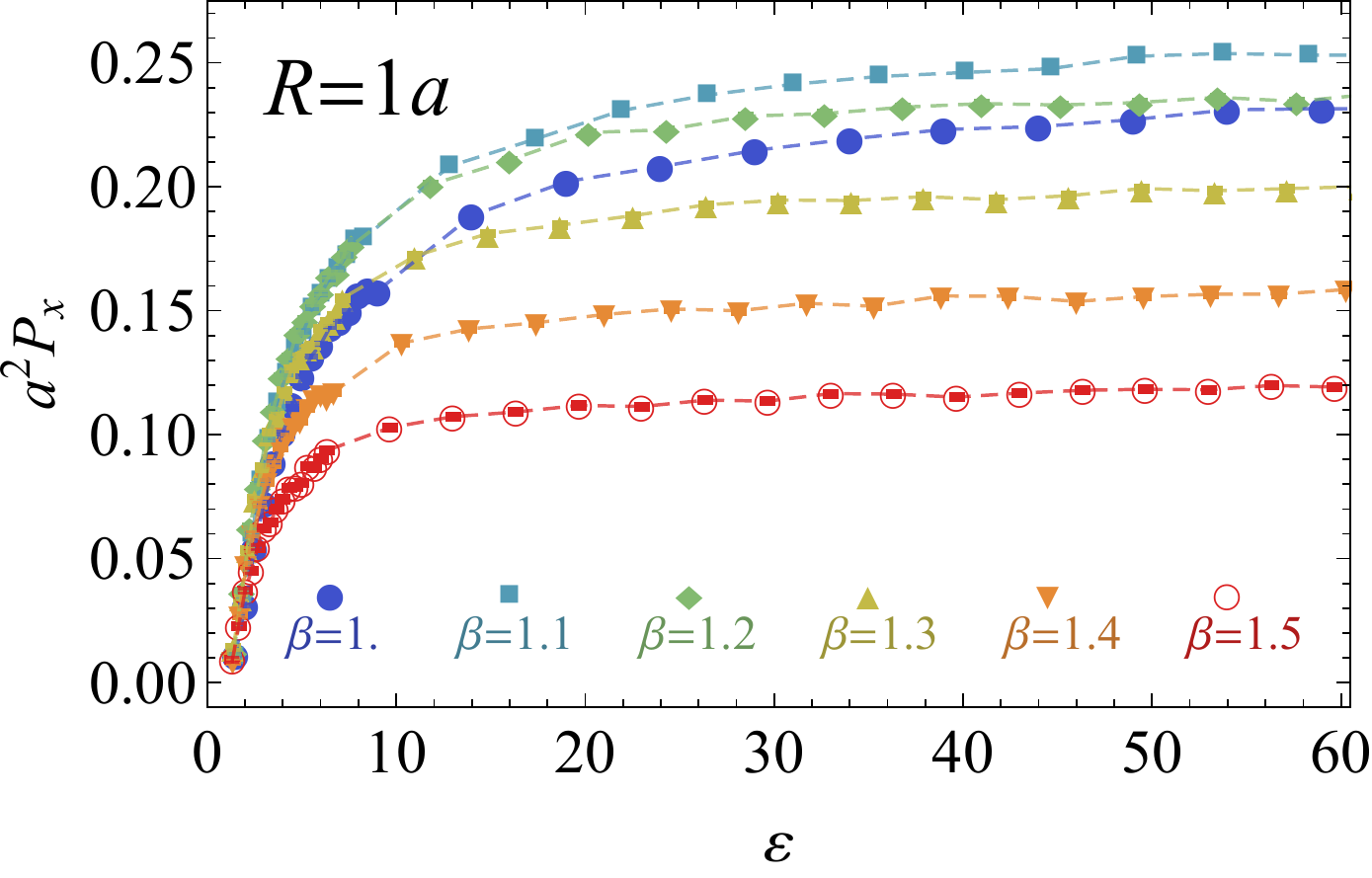}
\end{center}
\vskip -5mm 
\caption{Normal stress~\eq{eq:P:Cas:lat} vs. permittivity~$\varepsilon$ at minimal distance between the wires, $R = 1 a$. The lines are drawn to guide the eye.} 
\label{fig:T11:extrapolation}
\end{figure}

For the sake of completeness, we also show in Fig.~\ref{fig:T11:extrapolation} the normal stress~\eq{eq:P:Cas:lat} as the function of permittivity $\varepsilon$ (at $R = 1a$ the integrated normal stress is equal to the normal stress). The stress in our definition is the positive quantity which is a monotonically increasing function of~$\varepsilon$. It gets maximal values at $\beta \simeq 1.1$ which naturally, in view of Eq.~\eq{eq:F:R}, corresponds to a steepest slope of the Casimir potential shown Fig.~\ref{fig:Vcas:lattice}.

\section{Conclusions}

In our article we discussed nonperturbative features of the Casimir effect which arise due to the presence of dynamical topological defects. As an interesting example that can be treated both analytically and numerically, we considered the zero-point potential between two dielectric wires in a compact version of Abelian gauge theory in two spatial dimensions at zero temperature. The spectrum of this theory possesses topological defects, instanton-like monopoles, which are known to be responsible for a number of nonperturbative features in the model including the effect of mass gap generation. 

We calculated the zero-point Casimir potential in an analytical approach based on a duality transformation: the model is first transformed to a dual representation where a perturbative treatment is possible, and then the Casimir potential is evaluated directly in the dual model. We have shown that the dynamical monopoles make the Casimir potential short-ranged, Eqs.~\eq{eq:V:Cas:R:massive} and \eq{eq:f:x}, with the radius of interaction given by the inverse mass of the photon~\eq{eq:m:ph:rho}. The Casimir potential becomes dependent both on the monopole density and on the gauge coupling of the model. The calculation is valid in the dilute gas approximation where the number of monopoles in a unit Debye volume is small. 

Our numerical simulations complement the analytical calculations. Using the numerical method developed in Ref.~\cite{Chernodub:2016owp} we have shown that in the region of strong gauge coupling, where the monopole density is high, the Casimir potential is weak and it falls down very quickly with increase of the distance, Fig.~\ref{fig:Vcas:lattice}. Both these findings are consistent with the effect expected from a massive photon field (the photon becomes massive due to the nonperturbative mass generation caused by the presence of the dynamical monopoles). In the region of an intermediate gauge coupling the Casimir potential becomes enhanced by the monopoles. At a weak gauge coupling the density of monopoles becomes negligibly small and the potential reduces to the standard coupling-independent analytical result.

The behavior of zero-point potential at the intermediate coupling may be related to the effects of the wires on the monopole dynamics. 

Firstly, as we argued analytically and then found numerically, the monopole density is suppressed in between the wires. The monopole suppression is especially strong in the limit of large permittivity $\varepsilon$ of the wires. 

Secondly, the wires effectively squeeze the flux of the monopoles in between them thus making inter-monopole interactions two-dimensional. In two-dimensions the inter-monopole interaction is governed by a confining logarithmic Coulomb potential. Therefore the monopoles tend to form dilute monopole-antimonopole pairs (magnetic dipoles) which cannot generate the mass gap, and, consequently, cannot effectively inhibit the zero-point interactions between the wires. Thus, the formation of the magnetic dipoles in between the plates makes the Casimir potential long-ranged. Moreover, the  density of the monopoles outside the plates is much larger compared to the density in between the plates. This difference in monopole densities increases the external pressure on the wires and enhances the Casimir potential. At stronger gauge coupling the mentioned mechanism does not work due to large density of monopoles because the intermonopole distance become smaller than the typical size of the monopole-antimonopole pair so that the magnetic dipole picture is no more applicable. 

Summarizing, we have found that the dynamical topological defects modify nonperturbatively the zero-point Casimir interaction between dielectric bodies. In general, this finding may be relevant to a large class of effective infrared theories in condensed matter physics. 

\acknowledgments

The work was supported by the Federal Target Programme for Research and Development in Priority Areas of Development of the Russian Scientific and Technological Complex for 2014-2020 (the unique identifier of the Contract RFMEFI58415X0017, contract number 14.584.21.0017).  The numerical simulations were performed at the computing cluster Vostok-1 of Far Eastern Federal University.

\end{document}